\documentclass[10pt,journal,compsoc]{IEEEtran}
\usepackage{newtxtext}
\usepackage[utf8]{inputenc}
\usepackage[T1]{fontenc}
\usepackage{pgfplots}
\pgfplotsset{compat=1.18}

\usepackage{pifont}
\usepackage{xcolor}
\newcommand{\cmark}{\textcolor{green!50!black}{\ding{51}}}
\newcommand{\xmark}{\textcolor{red!70!black}{\ding{55}}}

\usepackage{amsmath,amssymb,booktabs,tabularx,array,multirow,graphicx,xurl,cite,listings,xcolor}

\lstdefinelanguage{GHAYaml}{
    sensitive=true,
    alsoletter={-_},
    morekeywords=[1]{
        name,on,jobs,steps,uses,run,if,
        runs-on,permissions,env,concurrency,
        group,cancel-in-progress
    },
    morekeywords=[2]{
        pull_request,pull_request_target,
        push,schedule,workflow_dispatch,
        write,read,true,false
    },
    morecomment=[l]{\#},
    morestring=[b]",
    morestring=[b]'
}

\lstdefinestyle{gha-motivation}{
    language=GHAYaml,
    basicstyle=\ttfamily\scriptsize,
    keywordstyle=[1]\color{blue!70!black}\bfseries,
    keywordstyle=[2]\color{violet!75!black},
    commentstyle=\color{gray!75!black}\itshape,
    stringstyle=\color{teal!65!black},
    numbers=none,
    numberstyle=\tiny\color{gray},
    numbersep=6pt,
    frame=single,
    rulecolor=\color{gray!40},
    backgroundcolor=\color{gray!4},
    columns=fullflexible,
    breaklines=true,
    breakatwhitespace=true,
    showstringspaces=false,
    captionpos=b,
    xleftmargin=0pt,
    framexleftmargin=4pt,
    aboveskip=6pt,
    belowskip=6pt
}
\usepackage{tikz}
\usepackage{tcolorbox}

\newtcolorbox{rqanswer}[1]{
    colback=gray!5,
    colframe=gray!55,
    colbacktitle=gray!12,
    coltitle=black,
    title={RQ$_{#1}$ Answer},
    fonttitle=\bfseries\footnotesize,
    fontupper=\footnotesize,
    sharp corners,
    boxrule=0.4pt,
    boxsep=1.5pt,
    left=1.5mm,
    right=1.5mm,
    top=1mm,
    bottom=1mm,
    before skip=5pt,
    after skip=7pt
}
\usetikzlibrary{positioning,arrows.meta,shapes.misc,calc}
\usepackage[colorlinks=true,allcolors=blue]{hyperref}
\newcommand{\jev}{\textsc{Jev~1.13}}
\newcommand{\gptdec}{\textsc{GPT Decisions}}
\newcommand{\luna}{\textsc{GPT-6 Luna}}
\newcommand{\tevone}{\textsc{Tev1~4B}}

\newcommand{\gemmafour}{\textsc{Gemma~4 E4B}}

\newcommand{\smell}[1]{\texttt{#1}}
\newcolumntype{Y}{>{\raggedright\arraybackslash}X}
\newcolumntype{C}{>{\centering\arraybackslash}X}
\lstdefinestyle{gha}{basicstyle=\ttfamily\scriptsize,columns=fullflexible,breaklines=true,breakatwhitespace=true,showstringspaces=false,frame=single,framesep=3pt,xleftmargin=0pt,xrightmargin=0pt}
\begin{document}

\title{Decision Models for Definition-Driven GitHub Actions Workflow Smell Detection}

\author{
    \IEEEauthorblockN{
        Rohit Gheyi\IEEEauthorrefmark{3}, Anderson Uchôa\IEEEauthorrefmark{4}, Wesley Assunção\IEEEauthorrefmark{1}, Francisco Sousa\IEEEauthorrefmark{3}, Baldoino Fonseca\IEEEauthorrefmark{2}, Márcio Ribeiro\IEEEauthorrefmark{2} \\
    }
    \IEEEauthorblockA{
        \IEEEauthorrefmark{3}Federal University of Campina Grande, Brazil\\Email: rohit@dsc.ufcg.edu.br and francisco.antonio.sousa@ccc.ufcg.edu.br\\
    }
    \IEEEauthorblockA{
        \IEEEauthorrefmark{4}Federal University of Ceará, Brazil\\Email: andersonuchoa@ufc.br\\          
    }
    \IEEEauthorblockA{
        \IEEEauthorrefmark{1}North Carolina State University, USA\\Email: wguezas@ncsu.edu\\          
    }       
    \IEEEauthorblockA{
        \IEEEauthorrefmark{2}Federal University of Alagoas, Brazil\\Email: baldoino@ic.ufal.br and marcio@ic.ufal.br\\
    } 
}

\maketitle

\begin{abstract}
GitHub Actions workflows automate continuous integration and delivery (CI/CD), and workflow smells are recurring configuration patterns that may weaken security, hinder maintenance, or waste resources. Extending static analyzers to new smells requires encoding constructs, applicability conditions, and exceptions in detection rules. We investigate definition-driven detection, in which models assess workflows against reusable natural-language smell definitions, focusing on decision-oriented models that return scores for bounded questions. We compare them with generative models and a static analyzer, and examine their score reliability, repeatability, and robustness. We evaluate three decision-oriented and two generative configurations on 1,710 manually validated workflow--smell pairs covering 19 smells, 903 workflows, and 559 repositories. Effectiveness depends strongly on the decision policy: \jev{} achieves 79.69\% micro-F1 at threshold 0.5 and 85.43\% with thresholds selected on separate repository folds, but global thresholds selected on the benchmark lower micro-F1 on a historical comparison set. Generative \luna{} achieves comparable effectiveness on the pairs it classifies but leaves 247 unresolved. Repeatability does not ensure robustness: two decision models agree across three identical-input runs yet change more than 3\% of their predictions after label-preserving transformations, despite nearly unchanged aggregate F1. On 46 reconstructed workflows, a static analyzer outperforms \jev{} (87.8\% versus 76.4\% micro-F1). With a median latency of 0.675 seconds per workflow and about US\$0.22 in API charges for the benchmark, \jev{} makes frequent model-assisted checks operationally plausible, but deployment requires validated decision policies, and effectiveness on new criteria remains to be evaluated.
\end{abstract}


\section{Introduction}
\label{sec:intro}

Continuous integration and delivery (CI/CD) pipelines automate software building, testing, and deployment~\cite{vassallo2020cdlinter}. In GitHub Actions, this automation is configured through YAML workflows that define event triggers, jobs, permissions, and external actions~\cite{decan2022use}. Although these constructs support flexible automation, their configuration can expose privileged operations, undermine reproducibility, or waste computational resources~\cite{koishybayev2022characterizing,muralee2023argus,bouzenia2024resource}. Workflow smells capture recurring patterns associated with such concerns, even when they do not constitute errors or cause execution failures~\cite{khatami2024catching}. Detecting them requires both localized checks, such as identifying mutable action references, and contextual assessments of conditions, execution scopes, and project requirements~\cite{khatami2024catching,uchoa2026gha}. Automated detection must therefore identify relevant problems while limiting false alarms, which add review effort and can discourage adoption~\cite{johnson2013static,sadowski2015tricorder}.

To automate these checks, static analyzers encode supported workflow properties as explicit detection rules or semantic analyses~\cite{khatami2024catching,freitas2024gash,actionlint,zizmor}. Extending their coverage requires translating additional criteria into analysis logic that accounts for relevant constructs, interactions across workflow scopes, and exceptions. Natural-language specifications offer a complementary approach~\cite{rian-agentes-test-smells-2026,melo2026reftestmodel,pedro-sbes-2025}: developers describe an undesirable pattern, its applicability conditions, and exceptions, and a model evaluates workflows against that description. New checks can thus be expressed without implementing dedicated detection rules, although the correctness of the model's interpretation requires validation. This approach builds on declarative smell detection in source code and Infrastructure as Code~\cite{moha2010decor,saavedra2022glitch} and language-model inspection of source code, infrastructure configurations, and GitHub Actions workflows~\cite{silva2024chatgpt,devito2025secllm,abrokwah2026compliant}. These precedents motivate investigating how reliably models apply workflow smell specifications, particularly when exceptions or unavailable project context affect the judgment.

These specifications can be evaluated through different model interfaces. Generative models can produce classifications with explanations, whereas decision-oriented models answer bounded questions, such as whether a workflow exhibits a specified smell, using labels or numerical scores~\cite{geuter2026inner}. Their focused output supports automated classification and adjustable thresholds without requiring generated explanations, motivating investigation of their suitability for frequent workflow analysis. However, numerical scores are not necessarily calibrated probabilities, and threshold choice affects classification results~\cite{guo2017calibration,lipton2014thresholding}. Allowing uncertain responses also leaves some cases unresolved, trading coverage for selective prediction~\cite{geifman2017selective}. Moreover, agreement across identical-input runs does not establish robustness to edits that preserve the assessed properties~\cite{atil2025nondeterminismdeterministicllmsettings,metamorphic-testing}. Although prior research has investigated LLM-based auditing of GitHub Actions workflows~\cite{abrokwah2026compliant}, less is known about the effectiveness of decision-oriented configurations under different threshold policies and the stability of their predictions under label-preserving workflow changes. These aspects warrant evaluation alongside detection effectiveness and inference cost.

In this work, we investigate \emph{definition-driven workflow smell detection}, in which models assess workflows against reusable natural-language specifications. Our goal is to determine whether this approach detects smells accurately, consistently, and at low inference cost, and how its results compare with specialized static analysis. We evaluate three decision-oriented configurations, comprising two proprietary models (\jev{} and \gptdec{}) and one open-weight model (\tevone{}), alongside two generative configurations (\luna{} and \gemmafour{}). We reuse a benchmark from a previous study involving some of the authors~\cite{uchoa2026gha}, containing 1,710 workflow--smell pairs whose detector-generated candidates were confirmed by manual annotation, covering 19 smells, 903 workflows, and 559 open-source repositories. We examine threshold selection, abstention, score reliability, and inference cost. For decision models, repeated runs assess consistency on identical inputs, while metamorphic testing~\cite{metamorphic-testing} uses five transformations designed and checked to preserve the assessed smell labels, allowing us to examine whether changes to workflow representation alter individual predictions. To compare decision models with static analysis, we separately use historical manual assessments from Khatami et al.~\cite{khatami2024catching} on 46 reconstructed workflows.

The results highlight three considerations for workflow smell detection. First, threshold selection improves effectiveness within the benchmark: \jev{} reaches 85.43\% micro-F1, narrowing its gap to \gptdec{} from 9.01 to 3.75 percentage points. However, applying global thresholds selected on this benchmark to the 46 reconstructed workflows reduces micro-F1, indicating limited transfer of these policies. Second, repeatability does not ensure stable diagnostics: two models agree across three identical-input runs yet change more than 3\% of their predictions after edits intended to preserve the assessed smells, despite little change in aggregate F1. Third, the static analyzer performs best on the reconstructed workflows, particularly for localized smells. Exploratory model-based filtering rejects some false alarms but also suppresses true positives, leaving its practical benefit to independent evaluation. Together with specification-based flexibility, \jev{}'s median request latency of 0.675 seconds and approximately US\$0.22 in benchmark API charges motivate prototyping additional checks; effectiveness on new criteria remains untested.
The article makes two empirical contributions:
\begin{itemize}
\item \textbf{An evaluation of definition-driven workflow smell detection.} We assess detection effectiveness and operational cost across decision-oriented and generative configurations, accounting for threshold selection and abstention, and compare decision models with a specialized static analyzer.
\item \textbf{An analysis of workflow smell prediction reliability.} Repeated inference and metamorphic testing assess whether decision-model predictions remain consistent across identical workflows and edits that preserve the assessed smells, revealing changes concealed by aggregate effectiveness.
\end{itemize}

Section~\ref{sec:motivation} presents the background and motivating example, and Section~\ref{sec:approach} describes the approach. Sections~\ref{sec:design} and~\ref{sec:results} present the study design and results. Sections~\ref{sec:discussion}--\ref{sec:conclusion} cover implications, threats to validity, related work, and conclusions.

\section{Background and Motivating Example}
\label{sec:motivation}

This section introduces GitHub Actions workflows and workflow smells, followed by a real-world example motivating definition-driven detection.

\subsection{GitHub Actions Workflows}

GitHub Actions automates software development activities through workflows defined in YAML files~\cite{decan2022use,github2026syntax}. A workflow specifies its triggering events, jobs, and steps. Common triggers include \texttt{push}, \texttt{pull\_request}, and \texttt{pull\_request\_target}~\cite{github2026syntax}. Jobs execute on designated runners and may invoke shell commands, reusable workflows, or external actions.

Several constructs influence security and execution behavior~\cite{github2026syntax,koishybayev2022characterizing,muralee2023argus}. The \texttt{permissions} field controls access granted to the GitHub authentication token, while \texttt{if} conditions determine whether jobs or steps execute. The \texttt{concurrency} construct controls overlapping executions and cancellation of superseded runs. These constructs underlie smells such as \smell{MissingTokenPermissions} and \smell{MissingConcurrencyPR} (Table~\ref{tab:catalog}). A smell indicates a potential concern; it does not by itself establish an exploitable vulnerability. 

\subsection{Workflow Smells and Static Detection}

Workflow smells are recurring configuration patterns that may negatively affect security, maintainability, or resource utilization without causing immediate execution failures~\cite{khatami2024catching, uchoa2026gha}. Examples include missing execution timeouts, mutable action versions, overly permissive authentication tokens, and inadequate concurrency controls. Localized smells include mutable runner labels; contextual smells require relating triggers, conditions, and execution behavior~\cite{khatami2024catching}. For example, assessing restrictions on privileged operations requires examining their triggering events and effective guards, concerns also studied in workflow security analysis~\cite{koishybayev2022characterizing,muralee2023argus}.

Static analyzers address different workflow properties. GASH focuses on workflow smells~\cite{freitas2024gash}, while GHAST and ARGUS examine potentially unsafe configurations and execution paths~\cite{benedetti2022automatic,muralee2023argus}. Tools such as actionlint and zizmor support workflow validation and security checking~\cite{actionlint,zizmor}. These analyzers encode supported properties as rules or semantic analyses. Extending their coverage requires corresponding changes to the analysis logic. Natural-language specifications provide an alternative way to express detection criteria, including applicability conditions and exceptions, while leaving their interpretation to a model. Whether this approach detects smells reliably or reduces the effort of introducing new checks requires empirical evaluation.

\subsection{Motivating Example}

Consider a historical version of the \texttt{.github/workflows/cifuzz.yml} workflow from \texttt{google/gson}, an open-source Java library for JSON serialization.\footnote{Recovered revision: \url{https://github.com/google/gson/blob/1bde0778ba15bfacb82787e144f7c6efed56d548/.github/workflows/cifuzz.yml}.} Listing~\ref{lst:motivation} shows how the workflow builds fuzzers, runs them for up to 600 seconds on pull requests, and uploads crash artifacts after failures.

\begin{lstlisting}[
    style=gha-motivation,
    morekeywords={
        name,on,jobs,steps,uses,run,if,id,with,
        runs-on,permissions,env,concurrency,
        group,cancel-in-progress,timeout-minutes,
        needs,strategy,matrix,outputs,secrets,environment
    },
    upquote=true,
    escapeinside={(*@}{@*)},
    breaklines=true,
    breakatwhitespace=false,
    breakautoindent=true,
    breakindent=0pt,
    postbreak=\mbox{\textcolor{gray}{$\hookrightarrow$}\space},
    columns=fullflexible,
    linewidth=\linewidth,
    captionpos=b,
caption={Excerpt from a historical \texttt{CIFuzz} workflow in \texttt{google/gson} exhibiting \smell{MissingConcurrencyPR} (action commit hash abbreviated).},
    label={lst:motivation}
]
name: CIFuzz
on: [pull_request]
jobs:
  Fuzzing:
    runs-on: ubuntu-latest
    steps:
    - name: Build Fuzzers
      id: build
      uses: google/oss-fuzz/infra/cifuzz/actions/build_fuzzers@master
      with:
        oss-fuzz-project-name: 'gson'
        dry-run: false
        language: jvm
    - name: Run Fuzzers
      uses: google/oss-fuzz/infra/cifuzz/actions/run_fuzzers@master
      with:
        oss-fuzz-project-name: 'gson'
        fuzz-seconds: 600
        dry-run: false
    - name: Upload Crash
      uses: actions/upload-artifact@6546280
      if: failure() && (*@steps@*).build.outcome == 'success'
      with:
        (*@name@*): artifacts
        path: ./out/artifacts
\end{lstlisting}

The workflow exhibits the \smell{MissingConcurrencyPR} smell: it responds to pull-request updates but defines no workflow-level or job-level concurrency policy to cancel superseded runs. Successive commits can therefore leave outdated fuzzing runs executing alongside newer ones, potentially wasting CI time and compute resources~\cite{github2026syntax}. The artifact-upload condition controls whether crash artifacts are uploaded; it does not prevent redundant executions. Khatami et al.\ manually labeled this workflow as exhibiting the smell~\cite{khatami2024catching}. Their study also documents historical fixes and favorable maintainer responses to proposed corrections.

Automating this check requires detection logic that accounts for event triggers, concurrency groups, cancellation settings, and their scope. Although this example makes the smell evident because no concurrency policy is defined, the presence of a \texttt{concurrency} field does not necessarily resolve it. Successive runs of the same pull request must share an appropriate concurrency group, with \texttt{cancel-in-progress} enabled to cancel superseded executions~\cite{github2026syntax}.

Natural-language specifications express the same criterion as a definition and detection guidance (Table~\ref{tab:smell-specification}). The guidance identifies the relevant configuration constructs and the conditions under which the smell should not be reported, so a model must apply these criteria to the workflow rather than merely check for the presence of a configuration field. Prior studies have used language models to detect workflow problems with generative prompting~\cite{uchoa2026gha,abrokwah2026compliant}, but whether decision models, which return scores for bounded questions without generating text, can apply such specifications remains unexamined. It is unknown whether their judgments are stable across repeated runs and edits that preserve the assessed smells, and how they compare with a specialized static analyzer. We investigate these questions in this study.

\section{Definition-Driven Detection}
\label{sec:approach}

Definition-driven detection separates what constitutes a smell from how a model is queried. A specification defines the criterion to be evaluated, an inference adapter submits it to a model, and a decision policy converts the model's output into a classification. This section describes the specifications and their conversion into predictions.

\subsection{Declarative Smell Specifications}
\label{sec:catalog}

Each specification contains a smell name, a natural-language definition of the undesirable configuration, and guidance on detection, applicability, and exceptions. A domain expert drafts these fields and may use an LLM to clarify their wording, retaining responsibility for the final criterion. The specification describes the property to check; instructions about response format are handled separately (Section~\ref{sec:inference}).
Table~\ref{tab:smell-specification} illustrates \smell{MissingConcurrencyPR}, introduced in Section~\ref{sec:motivation}. Its guidance identifies the settings relevant to determining whether superseded runs are canceled.

\begin{table}[t]
\centering
\caption{Illustrative specification of \smell{MissingConcurrencyPR}.}
\label{tab:smell-specification}
\footnotesize
\begin{tabularx}{\columnwidth}{@{}lX@{}}
\toprule
\textbf{Field} & \textbf{Content} \\
\midrule
Name & \smell{MissingConcurrencyPR} \\
\addlinespace
Definition &
Superseded pull-request runs are not automatically canceled. \\
\addlinespace
Guidance &
Inspect pull-request triggers and workflow- and job-level concurrency policies. Exclude configurations in which successive runs of the same pull request share an appropriate concurrency group with \texttt{cancel-in-progress} enabled, canceling the relevant superseded runs. \\
\bottomrule
\end{tabularx}
\end{table}

The smell catalog used in this study builds on the workflow smells investigated by Khatami et al.~\cite{khatami2024catching} and the operational definitions developed in a previous study~\cite{uchoa2026gha}. Table~\ref{tab:catalog} summarizes 19 smells covering security, maintainability, and resource efficiency. We distinguish predominantly localized criteria from those requiring broader workflow context. This grouping reflects the primary evidence needed for each criterion rather than a strict separation: even localized checks may involve applicability conditions or exceptions, while some contextual criteria depend on information unavailable in the workflow YAML.

\begin{table*}[t]
\centering
\caption{Evaluated workflow smells, grouped by the predominant scope of evidence required. Definitions are abbreviated; complete specifications, including guidance and exceptions, are provided in the replication materials (Data Availability).}
\label{tab:catalog}
\scriptsize
\begin{tabularx}{\textwidth}{@{}p{.275\textwidth}p{.085\textwidth}Y@{}}
\toprule
\textbf{Smell} & \textbf{Scope} & \textbf{Definition (summary)} \\
\midrule
\smell{MutableRunnerVersion} & Local & A job uses a mutable runner label, such as \texttt{ubuntu-latest}, instead of a fixed runner version. \\
\smell{SuboptimalCacheUsage} & Local & A workflow uses separate caching mechanisms where appropriate setup-action caching is available. \\
\smell{MissingTokenPermissions} & Local & The workflow lacks appropriately restricted permissions for the GitHub authentication token. \\
\smell{MissingUploadCondition} & Local & An artifact-upload operation lacks an applicable execution condition. \\
\smell{MissingJobTimeout} & Local & A locally defined job lacks an explicit execution timeout. \\
\smell{MissingPermExternalActions} & Local & External actions are executed without appropriately restricted token permissions. \\
\smell{UnpinnedActionVersion} & Local & A remote action uses a mutable version reference instead of a full commit SHA. \\
\smell{MultiCommandStep} & Local & A single execution step combines independent operations that could be separated. \\
\smell{UnpinnedDependencies} & Local & External packages or tools are installed without appropriately fixed versions. \\
\smell{UnsafeArtifactUploadFork} & Local & Artifact-upload operations can execute on forks without an effective guard. \\
\smell{SingleLangVersionCI} & Context & CI tests only one language or runtime version despite relevant compatibility requirements. \\
\smell{MissingConcurrencyPR} & Context & Superseded pull-request runs are not canceled through appropriate concurrency controls. \\
\smell{MissingPathFilters} & Context & CI executes for unrelated file changes without suitable path filtering. \\
\smell{MissingForkGuardPR} & Context & PR- or issue-oriented operations that should not execute on forks lack effective guards. \\
\smell{SingleOSExecution} & Context & Relevant testing executes on only one operating system without multi-OS coverage. \\
\smell{MissingForkGuardSchedule} & Context & Scheduled execution lacks appropriate repository or fork restrictions. \\
\smell{MissingConcurrencyBranch} & Context & Superseded branch runs are not canceled through appropriate concurrency controls. \\
\smell{MissingConcurrencyGlobal} & Context & Concurrent executions that should be serialized can overlap. \\
\smell{UnsafeDeployOnFork} & Context & Deployment operations may execute on forks without appropriate restrictions. \\
\bottomrule
\end{tabularx}
\end{table*}

A check can be refined by modifying its specification without implementing a new detection rule. For instance, its guidance can require concurrency groups to distinguish pull requests, rather than merely require a group to exist. This illustrates how detection criteria can be revised independently of the inference mechanism, but does not establish that models will correctly apply the revised criteria. We do not measure authoring effort or adaptation to evolving workflow syntax.

\subsection{Definition-Driven Inference}
\label{sec:inference}

A workflow--smell pair $(w,s)$ asks whether workflow $w$ exhibits smell $s$. An \emph{inference adapter} is the software component that formats the workflow YAML and smell specification for a model, submits the request, and extracts its response. For example, a decision adapter requests a score for the presence of \smell{MissingConcurrencyPR}; a generative adapter can request a structured verdict and explanation. Requests may batch several pairs, as specified in Section~\ref{sec:implementation}. Figure~\ref{fig:study-overview} illustrates this flow within the evaluation. The adapters preserve the intended smell criteria while accommodating model-specific input and output formats. Consequently, differences in predictions may reflect both model behavior and the inference configuration.

For a numerical interface, let $p_s(w)\in[0,1]$ denote the returned smell-presence score. A threshold $\tau$ converts it into a binary prediction:
\begin{equation}
\widehat{y}(w,s)=
\begin{cases}
1\;\text{(smell present)}, & p_s(w)\geq\tau,\\
0\;\text{(smell absent)}, & p_s(w)<\tau.
\end{cases}
\label{eq:decision}
\end{equation}

For discrete responses, the adapter reads the label directly. An explicit \texttt{UNCERTAIN} response is an abstention, not a negative prediction. A response missing the required label is malformed, while a request timeout is an inference failure; neither establishes smell absence. Reporting these outcomes separately distinguishes effectiveness on classified cases from coverage~\cite{geifman2017selective}. Numerical scores also require calibration assessment before being interpreted as probabilities of smell presence~\cite{guo2017calibration}.

\section{Study Design}
\label{sec:design}

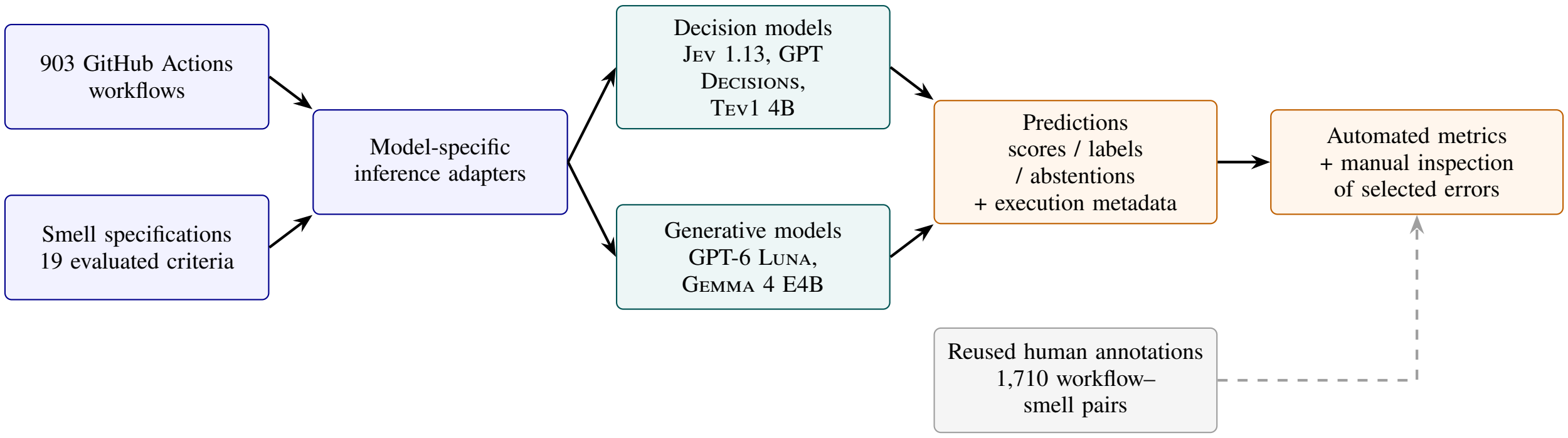
\begin{figure*}[t]
\centering
\begin{tikzpicture}[
    font=\scriptsize,
    box/.style={
        draw=blue!55!black,
        rounded corners=2pt,
        align=center,
        minimum height=11mm,
        inner sep=4pt,
        fill=blue!5
    },
    model/.style={
        box,
        draw=teal!65!black,
        fill=teal!7
    },
    result/.style={
        box,
        draw=orange!75!black,
        fill=orange!7
    },
    reference/.style={
        box,
        draw=gray!75,
        fill=gray!8
    },
    flow/.style={
        -{Stealth[length=2mm]},
        thick
    },
    referenceflow/.style={
        flow,
        dashed,
        draw=gray!75
    }
]

\node[box, text width=25mm] (wf)
    at (0,0.9)
    {903 GitHub Actions\\workflows};

\node[box, text width=25mm] (catalog)
    at (0,-0.9)
    {Smell specifications\\19 evaluated criteria};

\node[box, text width=24mm] (adapter)
    at (3.2,0)
    {Model-specific\\inference adapters};

\node[model, text width=26mm] (decision)
    at (6.5,1.0)
    {Decision models\\
     \jev{}, \gptdec{},\\
     \tevone{}};

\node[model, text width=26mm] (generative)
    at (6.5,-1.0)
    {Generative models\\
     \luna{},\\
     \gemmafour{}};

\node[result, text width=27mm] (predictions)
    at (9.9,0)
    {Predictions\\
     scores / labels / abstentions\\
     + execution metadata};

\node[result, text width=28mm] (evaluation)
    at (13.5,0)
    {Automated metrics\\
     + manual inspection\\
     of selected errors};

\node[reference, text width=27mm] (groundtruth)
    at (9.9,-2.3)
    {Reused human annotations\\
     1,710 workflow--smell pairs};

\draw[flow] (wf.east) -- (adapter.north west);
\draw[flow] (catalog.east) -- (adapter.south west);
\draw[flow] (adapter.east) -- (decision.west);
\draw[flow] (adapter.east) -- (generative.west);
\draw[flow] (decision.east) -- (predictions.north west);
\draw[flow] (generative.east) -- (predictions.south west);
\draw[flow] (predictions.east) -- (evaluation.west);

\draw[referenceflow]
    (groundtruth.east) -| (evaluation.south);

\end{tikzpicture}
\caption{Overview of the primary benchmark evaluation. Adapters submit workflows and smell specifications to five model configurations. Scripts compute metrics using recorded outputs and reused human annotations; manual inspection of selected errors supports interpretation.}
\label{fig:study-overview}
\end{figure*}

This section presents the research questions, dataset, model configurations, and evaluation procedures. Figure~\ref{fig:study-overview} summarizes the primary benchmark evaluation, conducted in October 2026.

\subsection{Goal and Research Questions}
\label{sec:rqs}

Following the Goal--Question--Metric template~\cite{basili1994gqm}, our goal is to \emph{analyze} definition-driven workflow smell
detection; \emph{for the purpose of} evaluation; \emph{with respect to} detection effectiveness, operational efficiency, and the repeatability and robustness of decision-model predictions, and agreement with historical manual assessments relative to a specialized static analyzer; \emph{from the point of view of} software engineering researchers and CI/CD tool builders; \emph{in the context of} manually annotated GitHub Actions workflows from open-source repositories. We address seven research questions (RQs).

\textbf{RQ$_1$ (Detection Effectiveness):} 
\emph{How accurately do decision-oriented and generative configurations detect GitHub Actions workflow smells?} A definition-driven detector is useful only if its judgments agree with expert assessments, and model choice is the first decision a tool builder faces. We compare precision, recall, and micro- and macro-F1, distinguishing effectiveness on classified pairs from classification coverage.

\textbf{RQ$_2$ (Per-Smell Performance and Complementary Errors):} \emph{How does effectiveness vary across smells, and to what extent do the configurations make complementary errors?} Aggregate scores can hide smells that a configuration systematically misses. We call errors \emph{complementary} when a pair misclassified by one configuration is correctly classified by another; such errors indicate whether configurations fail on different cases, and hence whether combining them could help, or share the same blind spots. We examine per-smell results, shared and exclusive true positives and errors, and representative false positives and false negatives.

\textbf{RQ$_3$ (Decision Thresholds and Abstention):} \emph{How do threshold selection and abstention affect detection effectiveness and classification coverage?} Decision interfaces return scores that must be converted into labels, and generative models may decline to answer, so the reported effectiveness depends on a decision policy rather than on the model alone. We evaluate alternative and cross-validated thresholds for decision models and examine uncertain responses from the selective generative configuration.

\textbf{RQ$_4$ (Operational Efficiency):} \emph{What token usage, latency, and API charges do the evaluated configurations incur?} Continuous feedback in CI pipelines or editors is practical only if detection is fast and inexpensive. We measure token usage and request latency, estimate hosted API charges, and distinguish these charges from unmeasured local execution costs.

\textbf{RQ$_5$ (Repeatability of Decision Models):} \emph{How consistent are decision-model predictions across repeated executions, and does majority voting improve effectiveness?} A detector whose warnings change between identical runs undermines developer trust and complicates evaluation. We compare labels and scores across three complete runs, assess majority-vote F1, and quantify the additional inference cost.

\textbf{RQ$_6$ (Robustness of Decision Models):} \emph{How do workflow transformations that preserve the assessed smell labels affect decision-model predictions?} Developers routinely make edits that do not affect a smell, such as adding comments or reformatting, and a robust detector should not change its warnings in response. Through metamorphic testing~\cite{metamorphic-testing-2}, we measure changes in aggregate effectiveness and individual predictions, using identical-input repetitions as a baseline for inference variability.

\textbf{RQ$_7$ (Comparison with Static Analysis):} \emph{How do decision models compare with a specialized static analyzer in detection effectiveness and error patterns?} Rule-based analyzers are the established alternative for workflow smell detection, so model-based detection must be positioned relative to them. We treat this comparison as a separate question rather than as part of RQ$_1$ because it requires a different reference standard: the main benchmark's candidate labels were generated by the analyzer under comparison (Section~\ref{sec:dataset}). We therefore score both approaches against manual assessments from the analyzer's original evaluation (Section~\ref{sec:external-design}), and compare localized and contextual smells and their complementary errors.

Decision models are the primary focus of the study. \textbf{RQ$_1$}--\textbf{RQ$_4$} compare them with generative configurations, which provide a reference for effectiveness, decision policies, and operational cost; \textbf{RQ$_5$} and \textbf{RQ$_6$} examine the repeatability and robustness of decision models; and \textbf{RQ$_7$} compares them with a specialized static analyzer. Specification-authoring effort and generalization to previously unseen smell definitions are outside the scope of the study.

\subsection{Dataset}
\label{sec:dataset}

We reuse a benchmark developed in an existing study of GitHub Actions workflow smells~\cite{uchoa2026gha}. The existing study collected workflows from open-source repositories and constructed manually validated positive and negative examples for 19 smell categories. Candidate examples were obtained by running Khatami et al.'s rule-based detector~\cite{khatami2024catching} and were subsequently reviewed by human annotators; the exception is the semi-synthetic \smell{UnsafeDeployOnFork} examples described next. We retained only candidates whose detector output was confirmed by the annotators; the benchmark therefore excludes cases of detector--annotator disagreement and evaluates agreement with confirmed detector outcomes. According to the existing study's annotation protocol, six annotators worked in three pairs and resolved disagreements through discussion, with a pre-consensus Cohen's $\kappa$ of 0.848. The quantitative evaluation covers the 19 smells with reference annotations. Three additional stylistic criteria in the catalog (\smell{AddCommentsToWorkflows}, \smell{CorrectIndentation}, and \smell{NameRunSteps}) lack annotations and are not scored.

The benchmark contains 1,710 annotated workflow--smell pairs from 903 workflows and 559 repositories, with 855 positive and 855 negative labels (Table~\ref{tab:dataset}). A workflow may appear in several pairs because it can be assessed for multiple smells; pairs without an annotation provide no evidence of smell presence or absence and are excluded from scoring. Of the 903 workflows, 858 are distinct natural workflow files and 45 are semi-synthetic variants of natural workflows also included in the benchmark, which the existing study introduced because naturally occurring positive \smell{UnsafeDeployOnFork} examples were scarce. These modifications were generated by an LLM (Claude~Opus~4.8) from real workflows manually labeled negative for this smell, restricted to the target-smell construct, manually validated, and marked as such in the dataset.

\begin{table}[t]
\centering
\caption{Characteristics of the reused workflow--smell benchmark. Languages are the primary repository languages reported by GitHub.}
\label{tab:dataset}
\footnotesize
\begin{tabular}{@{}lr@{}}
\toprule
\textbf{Characteristic} & \textbf{Value} \\
\midrule
Workflow instances & 903 \\
GitHub repositories & 559 \\
Evaluated smell categories & 19 \\
Annotated workflow--smell pairs & 1,710 \\
Positive / negative pairs & 855 / 855 \\
Natural / semi-synthetic workflows & 858 / 45 \\
Workflows with 0 / 1 / 2 / 3 positive annotations & 100 / 756 / 42 / 5 \\
Workflow size in LOC (min / median / max) & 7 / 45 / 207 \\
Repository languages & 27 \\
\bottomrule
\end{tabular}
\end{table}

Measured by the primary language that GitHub reports for each repository, the most represented languages are Python (142 workflows from 77 repositories), Go (135/92), TypeScript (112/87), Java (108/74), Rust (69/38), C++ (47/36), Shell (46/11), and C (45/31); the remaining 199 workflows from 113 repositories span 19 further languages. This diversity covers different workflow structures and automation tasks, although the benchmark is not a random sample of GitHub Actions workflows.

\subsection{Model Selection and Implementation}
\label{sec:implementation}

We selected five configurations spanning decision-oriented and generative inference, proprietary and open-weight models, and hosted and local deployment (Table~\ref{tab:models}). For decision-oriented inference, we selected \jev{}~\cite{typesafe2026api}, the decision model examined in several recent independent studies~\cite{geuter2026inner,yang2026limits,xu2026jevout}; \gptdec{}~\cite{openai2026decisions}, the decision interface of a second major provider; and \tevone{}~\cite{together2026tev1}, an open-weight decision model that can run locally. For generative inference, we selected a lightweight proprietary model, \luna{}, and an open-weight model, \gemmafour{}~\cite{google2026gemma4}, whose effective parameter count is similar to that of \tevone{}, so that each inference type is represented by one hosted and one local model.

\begin{table*}[t]
\centering
\caption{Primary configurations evaluated in October 2026. Decision models were executed three times and generative models once; primary effectiveness uses the first run.}
\label{tab:models}
\scriptsize
\begin{tabularx}{\textwidth}{@{}p{.14\textwidth}p{.13\textwidth}p{.21\textwidth}Y@{}}
\toprule
\textbf{Configuration} & \textbf{Inference type} & \textbf{Model / deployment} & \textbf{Main settings} \\
\midrule
\jev{} & Decision & \texttt{jev-1.13.0}; TypeSafe API & One question per smell, 22 questions per workflow request; numerical scores; threshold 0.5. \\
\gptdec{} & Decision & \texttt{gpt-6-luna}; OpenAI Decisions API & One question per smell, batched by workflow; numerical scores; threshold 0.5. \\
\tevone{} & Decision & \texttt{tev1-16k:4b}; Ollama & One question per smell, 22 questions per workflow request; temperature 0; 16K context; threshold 0.5. \\
\luna{} & Generative & \texttt{gpt-6-luna}; OpenAI Responses API & One request per annotated pair; low reasoning effort; 600 maximum output tokens; explanations. \\
\gemmafour{} & Generative & \texttt{gemma4:e4b}; Ollama & One request per workflow with 19 boolean judgments; temperature 0; 16K context; 1,024 maximum output tokens; thinking disabled. \\
\bottomrule
\end{tabularx}
\end{table*}

\noindent\textbf{Decision-oriented configurations.} We accessed \jev{} through TypeSafe's System One API~\cite{typesafe2026api}, \gptdec{} through OpenAI's Decisions API~\cite{openai2026decisions}, and \tevone{} locally through Ollama~\cite{ollama2026}. A decision model receives an input and a \emph{question}, a closed yes/no formulation of one smell specification (Section~\ref{sec:catalog}), and returns a score for the answer \emph{yes}; the replication package lists all questions. \tevone{} (released as \tevone{}-experimental; Ollama tag \texttt{tev1-16k:4b}) is queried through the same System One question schema as \jev{}; its score is the probability returned for each binary question. Each configuration received one question per catalog smell, with all 22 questions for a workflow submitted in one request; only pairs annotated for the 19 evaluated smells contributed to effectiveness metrics. Returned scores were converted into binary labels with the primary threshold of 0.5 (Equation~\ref{eq:decision}). \tevone{} is a decision-oriented fine-tune of Qwen3.5-4B~\cite{together2026tev1}. Its default 2,048-token runtime context was insufficient for some prompts, so we increased it to 16,384 tokens without changing the weights.

\noindent\textbf{Generative configurations.} \luna{} evaluated each annotated workflow--smell pair separately, returning \texttt{YES}, \texttt{NO}, or \texttt{UNCERTAIN}, a two- or three-sentence explanation, and supporting YAML evidence. It used the same model identifier as \gptdec{}, accessed through the Responses API; a shared identifier does not establish identical inference behavior across the two interfaces. The local \gemmafour{} configuration~\cite{google2026gemma4} received one workflow per request and returned structured boolean judgments for all 19 evaluated smells. Its exact installed model digest and quantization identifier were not retained. All 903 raw responses contain the 19 required boolean fields and report normal completion.

\noindent\textbf{Execution and comparability.} Local models ran through Ollama~\cite{ollama2026} on an Apple Mac Mini M4 Pro with 64\,GB of unified memory. For every request, we recorded the response, inference status, available token usage, and request latency. Adapters used the same intended smell criteria while accommodating each interface's input and output requirements. Because training, batching, prompting, and deployment differ, the results characterize complete configurations rather than isolating the effect of the inference interface.

\subsection{Evaluation Procedure and Metrics}
\label{sec:evaluation}

For each workflow, an adapter submits its YAML and the relevant smell specifications and maps the responses to annotated workflow--smell pairs. Adapters differ in request formatting and batching but target the same operational criteria; the replication package contains the complete specifications and model-specific prompts. Reference labels are never included in inference requests. They are used for scoring and, in cross-validation, for threshold selection on folds separate from those being evaluated. Only annotated pairs are scored; predictions for other smells are not treated as errors. Scripts compute all metrics from the recorded predictions and reference labels. We additionally inspect selected workflows and responses to interpret errors; these examples are illustrative, not a systematic qualitative coding study.

\noindent\textbf{Detection effectiveness (RQ$_1$).}
We count true positives (TP), false positives (FP), false negatives (FN), and true negatives (TN), with the smell-present label as the positive class, and compute precision, recall, and F1~\cite{manning2008ir}:
\begin{equation}
P=\frac{TP}{TP+FP},
\qquad
R=\frac{TP}{TP+FN},
\qquad
F_1=\frac{2PR}{P+R}.
\label{eq:metrics}
\end{equation}
Micro-averaged metrics pool all annotated pairs; macro-F1 averages the per-smell F1 values of the 19 smells. Decision interfaces use $\tau=0.5$ in the primary evaluation, whereas generative configurations return discrete labels. We compare micro-F1 between configurations using repository-cluster bootstrap confidence intervals~\cite{field2007bootstrapping} and paired permutation tests~\cite{good2005permutation}, with Holm correction~\cite{holm1979simple} for the six pairwise comparisons. Resampling repositories rather than pairs accounts for dependencies among pairs from the same project.

\noindent\textbf{Complementary errors (RQ$_2$).}
We compare the four full-coverage configurations on identical annotated pairs and identify the true positives and errors that are shared by all configurations or exclusive to one. Per-smell results and inspection of selected workflows help interpret these differences. Combinations of predictions, such as OR and AND rules, are evaluated descriptively; no ensemble is trained or independently validated.

\noindent\textbf{Decision thresholds and abstention (RQ$_3$).}
We inspect thresholds of 0.2, 0.3, 0.4, 0.5, and 0.6. These results are exploratory because the thresholds are inspected on the evaluation data. To separate threshold selection from evaluation, we also divide repositories into five folds; for each held-out fold, we select global or per-smell thresholds on the other four folds from 0.05 to 0.95 in steps of 0.01, and we pool the predictions of the five held-out folds. We assess discrimination with AUROC and average precision, and score reliability with the Brier score~\cite{brier1950verification} and expected calibration error (ECE) over ten equal-width bins~\cite{guo2017calibration}. For \luna{}, \texttt{UNCERTAIN} responses and failed requests are reported separately from binary predictions, distinguishing coverage from effectiveness on classified pairs~\cite{geifman2017selective}.

\noindent\textbf{Operational efficiency (RQ$_4$).}
We record token usage, request latency, successful and failed requests, and the usage metadata reported by each provider. Hosted API charges are estimated from the recorded usage and the prices in effect during the experiments~\cite{typesafe2026models,openai2026decisions}, including cached-token categories when reported. Local models incur no metered API charges; electricity, hardware, and maintenance costs were not measured. Token counts depend on tokenizers and reporting conventions.

\noindent\textbf{Repeatability (RQ$_5$).}
We compare the predictions for the same 1,710 annotated pairs across three complete runs of each decision configuration at threshold 0.5, reporting label agreement across all three runs, the spread of accuracy, and micro-F1 under majority voting. Generative configurations were not repeated under an unchanged protocol. The robustness evaluation for \textbf{RQ$_6$} and the comparison with static analysis for \textbf{RQ$_7$} are described in Sections~\ref{sec:metamorphic-design} and~\ref{sec:external-design}.

\subsection{Metamorphic Robustness Evaluation}
\label{sec:metamorphic-design}

To investigate \textbf{RQ$_6$}, we use metamorphic testing, which addresses the absence of a test oracle by checking relations that must hold between the outputs for related inputs~\cite{segura2016survey,metamorphic-testing-2}. We ask whether decision-model predictions remain unchanged after edits that preserve the assessed smell labels, and compare the resulting changes with those observed between repeated runs on identical inputs to separate transformation sensitivity from ordinary inference variability.

\noindent\textbf{Metamorphic relation.} For each original workflow $w$, we construct one transformed counterpart $T(w)$ and require $y_s(T(w))=y_s(w)$ for every assessed smell $s$, where $y_s$ is the reference label. Preserving execution behavior is not sufficient for this relation: explicitly adding \texttt{timeout-minutes: 360} leaves execution unchanged, because 360 minutes is the platform default, but turns a \smell{MissingJobTimeout} positive into a negative. Conversely, preserving the assessed labels does not establish complete runtime equivalence. The operators therefore never modify the assessed properties of \texttt{run} commands, \texttt{uses} references, triggers, permissions, timeouts, concurrency settings, runner specifications, or matrices; they never make default values explicit and never reorder step sequences.

\begin{table}[t]
\centering
\caption{Metamorphic operators and allocation. \emph{Elig.}: workflows
for which the operator produced a candidate that passed the preservation
checks; \emph{Wf.}: workflows assigned to the operator; \emph{Pairs}:
assessed positive/negative pairs in those workflows.}
\label{tab:metamorphic-operators}
\footnotesize
\setlength{\tabcolsep}{3pt}
\begin{tabularx}{\columnwidth}{@{}lXrrr@{}}
\toprule
\textbf{Op.} & \textbf{Transformation} & \textbf{Elig.} & \textbf{Wf.} & \textbf{Pairs} \\
\midrule
T1 & Insert neutral full-line comments & 903 & 182 & 170/168 \\
T2 & Change indentation and blank lines & 902 & 181 & 174/177 \\
T3 & Reorder mapping keys and jobs & 886 & 181 & 170/173 \\
T4 & Consistently rename job identifiers & 900 & 182 & 171/170 \\
T5 & Rewrite equivalent \texttt{if} expression syntax & 386 & 177 & 170/167 \\
\bottomrule
\end{tabularx}
\end{table}

\noindent\textbf{Operator design.} We designed the five operators in Table~\ref{tab:metamorphic-operators} for this study. They adapt to YAML workflows the semantics-preserving transformations commonly used to test machine-learning models of code, such as inserting comments, reformatting, and consistently renaming identifiers~\cite{metamorphic-testing}, and add two transformations specific to GitHub Actions: reordering mapping keys, to which the platform assigns no meaning, and the two equivalent syntaxes for conditions documented in the workflow syntax reference~\cite{github2026syntax}. Each operator alters a surface property that is irrelevant to every smell definition in the catalog:
\begin{itemize}
\item \emph{T1} inserts one to six full-line comments, approximately one per ten content lines, whose generic wording avoids every concept in the smell catalog.
\item \emph{T2} switches the indentation width between two and four spaces and separates top-level keys and jobs by blank lines.
\item \emph{T3} permutes mapping entries at the workflow level, among jobs, and within each job; sequences such as \texttt{steps} are never reordered.
\item \emph{T4} appends a suffix to every job identifier and rewrites the corresponding \texttt{needs} entries and \texttt{needs.}\emph{id}/\texttt{jobs.}\emph{id} references in expressions, preserving the dependency graph.
\item \emph{T5} adds or removes the \texttt{\$\{\{ \}\}} delimiters around standalone job- and step-level conditions, which GitHub Actions evaluates as expressions in either form~\cite{github2026syntax}; conditions that mix literal text and interpolation are left untouched, and unbraced conditions beginning with \texttt{!} are quoted as YAML requires.
\end{itemize}

\noindent\textbf{Preservation checks.} Every candidate was parsed and compared with its original under an operator-specific normalization: exact parsed-tree equality for T1 and T2, mapping equality irrespective of order for T3, equality after applying the same renaming to the original for T4, and equality after normalizing condition delimiters for T5. When an operator's candidate failed these checks, or the operator was not applicable (for example, T5 on a workflow without conditions), that operator was marked ineligible for the workflow and was not used for it; the \emph{Elig.} column of Table~\ref{tab:metamorphic-operators} reports the resulting counts. Every workflow had at least one eligible operator, because T1 was eligible for all 903, so every workflow was evaluated in transformed form. An independent verification pass re-checked all 903 emitted workflows, including that no protected construct changed and no comment was removed, and reported no violations. These checks establish that the configuration properties used by the smell definitions are preserved, not complete runtime equivalence: T4, for instance, changes the value of \texttt{github.job}, which no evaluated criterion inspects but which other steps could reference.

\noindent\textbf{Balanced allocation.} Each workflow received exactly one operator, chosen among its eligible operators by a seeded greedy procedure followed by local search. The allocation balanced the number of workflows per operator, the number of positive and negative pairs for each smell, and workflow origin (natural or semi-synthetic); the seed determines only the processing order and tie-breaking. The resulting groups contain 177 to 182 workflows and 337 to 351 pairs (Table~\ref{tab:metamorphic-operators}). In 34 of the 38 smell--label cells, the operator groups differ by at most two pairs, and in two further cells by four. The remaining two cells reflect a structural constraint: T5 requires a condition, whereas positive \smell{MissingForkGuardSchedule} and \smell{MissingConcurrencyPR} instances often lack one, so T5 received only one and six of their positive pairs, respectively. We therefore exclude T5 for these two smells from per-operator comparisons. Because eligibility differs across operators, the groups may still differ in workflow characteristics.

\noindent\textbf{Protocol and equivalence criterion.} Workflow identifiers, reference labels, and the 1,710 assessed pairs were preserved; only the YAML was transformed. \jev{}~\cite{typesafe2026api}, \gptdec{}~\cite{openai2026decisions}, and \tevone{}~\cite{together2026tev1} were run on the transformed workflows with their original configurations (Table~\ref{tab:models}) at threshold 0.5, scoring only annotated pairs. We compare original and transformed micro-F1 and, using the paired predictions, count changed binary labels (\emph{flips}), which we interpret against each model's flip rate between identical runs (\textbf{RQ$_5$}). To assess whether transformations leave aggregate effectiveness practically unchanged, we use an exploratory equivalence criterion: the micro-F1 difference between transformed and original workflows is considered negligible if its 90\% paired repository-cluster bootstrap confidence interval (10{,}000 resamples) lies entirely within $\pm2$ percentage points. The margin is a convention rather than a deployment-specific tolerance. Because equal aggregate F1 can hide compensating changes in individual predictions, we also report label flips.

\subsection{Comparison with a Static Analyzer}
\label{sec:external-design}

\noindent\textbf{Reference standard.} To investigate \textbf{RQ$_7$}, we compare the decision models with the rule-based detector released by Khatami et al.~\cite{khatami2024catching}. Both the detector and the models are scored against the same reference: the manual assessments that Khatami et al.\ produced when they evaluated their detector, not the labels of our main benchmark. A comparison on the main benchmark would be uninformative, because its positive and negative candidate labels were generated by this detector and then reviewed (Section~\ref{sec:dataset}): the released detector reproduces the reference label for 1,709 of the 1,710 annotated pairs, the remaining pair causing a parser error. 
Khatami et al.'s manual assessments are independent of our benchmark, but they were produced while evaluating the same detector, a limitation we discuss in Section~\ref{sec:threats}. Other GitHub Actions analyzers, such as GASH~\cite{freitas2024gash}, actionlint~\cite{actionlint}, and zizmor~\cite{zizmor}, implement related checks with different scopes and exceptions~\cite{fares2026unpacking}; comparing them would require aligning each check with our definitions, which we leave to future work.

\noindent\textbf{Reconstructing the evaluation set.} Khatami et al.'s replication package reports detector outputs and manual assessments for 115 workflows and 22 smells, of which we consider the 19 shared with our catalog (Table~\ref{tab:catalog}). The package identifies repositories and workflow paths, but not the analyzed commits, and most workflows have since changed. For each workflow, we retrieved the revisions of the file between January 2023 and the parent of the pull request that the original authors later submitted to the repository, ran the released detector on each revision, and selected the latest revision whose output matched the recorded detector output. We assigned each column to a smell by maximizing the agreement between recorded and recomputed detector outputs across workflows (agreement of at least 0.83 for every assigned column). We disabled the indentation check, which lies outside our catalog and is incompatible with the current \texttt{yamllint} configuration. Both revision selection and column mapping relied only on recorded detector outputs; the manual assessments were not consulted in either step.

For 46 of the 115 workflows, we recovered a revision whose detector output exactly matched the recorded output. For another 35, the closest revision differed on one or two smells; for 17, on three or more; and 17 workflow files could not be recovered. We restrict the comparison to the 46 exact matches: for the others, a manual assessment may refer to a configuration that is no longer present in the YAML given to the models, so scoring them would confound detection errors with reconstruction errors. The 46 workflows comprise 874 annotated workflow--smell pairs from 23 repositories, 275 of them positive (31.5\%). None of them occurs in the main benchmark, although two belong to a repository represented there. Because recovery was conditioned on the detector's own output, the subset may favor the detector, and the results are conditional on successful recovery.

\noindent\textbf{Protocol.} We evaluated \jev{}~\cite{typesafe2026api}, \gptdec{}~\cite{openai2026decisions}, and \tevone{}~\cite{together2026tev1} with their main-study configurations at threshold 0.5, and at thresholds selected on the main benchmark without external labels. The hosted models were scored on all 874 pairs. \tevone{} failed on one workflow, whose request exceeded the 64\,KB limit of the local Ollama server, and was scored on the remaining 855 pairs; its comparisons use the static detector's predictions on those same pairs, on which the detector reaches 88.2\% micro-F1. The failure is excluded from conditional effectiveness rather than treated as a negative prediction. Paired micro-F1 differences are reported with 95\% confidence intervals from 10{,}000 repository-cluster bootstrap resamples, separately for all smells and for the localized and contextual groups of Table~\ref{tab:catalog}.

\section{Study Results}
\label{sec:results}

This section presents the results for each research question.

\subsection{Detection Effectiveness (\textbf{RQ$_1$})}
\label{sec:rq1-results}

We first assess how effectively each configuration detects the 19 smells when used as released, without tuning its decision policy. Table~\ref{tab:performance} reports results at each configuration's default operating point. Four configurations return a binary label for all 1,710 pairs, whereas \luna{} classifies 1,463, abstains on 245, and fails on two requests; its effectiveness is therefore computed on classified pairs, with coverage reported separately.

\begin{table}[ht!]
\centering
\caption{Detection effectiveness on the annotated benchmark. Precision (P), recall (R), micro-F1, macro-F1, and scored coverage (\%).}
\label{tab:performance}
\footnotesize
\setlength{\tabcolsep}{5pt}
\resizebox{\linewidth}{!}{
\begin{tabular}{@{}lrrrrrrrrrr@{}}
\toprule
\textbf{Model} &
\textbf{Scored} &
\textbf{Scored coverage} &
\textbf{TP} &
\textbf{FP} &
\textbf{FN} &
\textbf{TN} &
\textbf{P} &
\textbf{R} &
\textbf{Micro-F1} &
\textbf{Macro-F1} \\
\midrule
\multicolumn{11}{@{}l}{\emph{Decision-oriented models}} \\
\jev{} & 1,710 & 100.00 & 620 & 81 & 235 & 774 & 88.45 & 72.51 & \textbf{79.69} & 75.40 \\
\gptdec{} & 1,710 & 100.00 & 505 & 69 & 350 & 786 & 87.98 & 59.06 & 70.68 & 58.69 \\
\tevone{} & 1,710 & 100.00 & 352 & 184 & 503 & 671 & 65.67 & 41.17 & 50.61 & 45.76 \\
\midrule
\multicolumn{11}{@{}l}{\emph{Generative models}} \\
\gemmafour{} & 1,710 & 100.00 & 152 & 120 & 703 & 735 & 55.88 & 17.78 & 26.97 & 18.54 \\
\luna{} & 1,463 & 85.56 & 573 & 111 & 166 & 613 & 83.77 & 77.54 & 80.53 & 74.93 \\
\bottomrule
\end{tabular}
}
\end{table}

\textbf{Finding 1: \jev{} achieves the highest detection effectiveness among full-coverage configurations, primarily through higher recall.}
Among configurations scored on all pairs, \jev{} achieves the highest micro-F1, exceeding \gptdec{} by 9.01 points. Their precision is similar (88.45\% versus 87.98\%), but \jev{} recovers more positives: recall is higher by 13.45 points. Low recall limits both local configurations. At this operating point, their F1 falls below the 66.67\% obtained by always predicting YES on this balanced benchmark.

\textit{Statistical comparisons.}
We compare the four full-coverage configurations at their default operating points. Confidence intervals use 10,000 paired repository-cluster bootstrap resamples, and two-sided tests use 20,000 permutations that swap predictions within repositories~\cite{efron1993bootstrap,davison1997bootstrap,field2007bootstrapping,good2005permutation}. Holm correction covers six comparisons~\cite{holm1979simple}. Positive differences in Table~\ref{tab:statistical-comparison} favor the first model. This accounts for within-repository dependencies; \luna{} is excluded because its scored subset differs. Table~\ref{tab:statistical-comparison} shows statistically significant differences in all six comparisons: every confidence interval excludes zero, and all Holm-adjusted $p$-values are below 0.001. Together, the comparisons support the micro-F1 ordering \jev{}, \gptdec{}, \tevone{}, and \gemmafour{} at the evaluated operating points.

\begin{table}[ht!]
\centering
\caption{Paired micro-F1 differences at threshold 0.5 (percentage points). CIs use repository-cluster bootstrap; $p$-values are Holm-adjusted.
}
\label{tab:statistical-comparison}
\footnotesize
\setlength{\tabcolsep}{8pt}
\resizebox{\linewidth}{!}{
\begin{tabular}{@{}lrrc@{}}
\toprule
\textbf{Comparison} &
\textbf{$\Delta$F1 (p.p.)} &
\textbf{95\% CI (p.p.)} &
\textbf{Holm-adjusted $p$} \\
\midrule
\jev{} vs. \gptdec{} & +9.01 & [6.14, 12.14] & $<0.001$ \\
\jev{} vs. \tevone{} & +29.08 & [24.06, 33.95] & $<0.001$ \\
\jev{} vs. \gemmafour{} & +52.72 & [48.24, 57.06] & $<0.001$ \\
\gptdec{} vs. \tevone{} & +20.07 & [13.78, 26.45] & $<0.001$ \\
\gptdec{} vs. \gemmafour{} & +43.71 & [38.14, 49.18] & $<0.001$ \\
\tevone{} vs. \gemmafour{} & +23.64 & [18.66, 28.72] & $<0.001$ \\
\bottomrule
\end{tabular}
}
\end{table}

\textbf{Finding 2: \luna{} achieves competitive detection effectiveness on classified pairs, but incomplete coverage limits direct comparisons.} On the 1,463 pairs classified by \luna{}, \jev{} achieves 78.93\% micro-F1 versus 80.53\% for \luna{}. The 1.60-point difference in favor of \luna{} has a 95\% CI of [$-0.83$, 3.98], providing insufficient evidence of an advantage. On the 247 pairs unresolved by \luna{}, \jev{} achieves 84.51\% F1, indicating that those cases are not uniformly difficult across configurations. Consequently, \luna{}'s higher conditional micro-F1 does not establish superior effectiveness over the full benchmark, since its performance on unresolved pairs remains unknown.
These comparisons describe the default operating points; Section~\ref{sec:rq3-results} examines how threshold selection changes the gaps.

\begin{rqanswer}{1}
At the default threshold, \jev{} is the most effective full-coverage configuration (79.69\% micro-F1); \luna{}'s comparable conditional F1 does not establish an advantage because it leaves 14.4\% of pairs unresolved.
\end{rqanswer}

\subsection{Per-Smell Performance and Complementarity (\textbf{RQ$_2$})}
\label{sec:rq2-results}

Table~\ref{tab:smelldiff} reports F1 for all 19 evaluated smells. The four full-coverage configurations are scored on identical annotated pairs; \luna{} is scored only on pairs receiving a binary judgment, so its coverage varies by smell. The benchmark contains 855 positive and 855 negative labels, with unequal support across smells.

\begin{table*}[t]
\centering
\caption{Per-smell F1 (\%) at the default operating point for 19 smells. \luna{} scores exclude uncertain responses and failures.}
\label{tab:smelldiff}
\footnotesize
\setlength{\tabcolsep}{6pt}
\begin{tabular}{@{}lrrrrr@{}}
\toprule
\textbf{Smell} &
\textbf{\jev{}} &
\textbf{\gptdec{}} &
\textbf{\tevone{}} &
\textbf{\gemmafour{}} &
\textbf{\luna{}} \\
\midrule
\smell{MissingConcurrencyBranch} & 99.01 & 90.91 & 63.16 & 14.81 & 72.90 \\
\smell{MissingConcurrencyGlobal} & 63.16 & 3.92 & 0.00 & 7.41 & 70.59 \\
\smell{MissingConcurrencyPR} & 100.00 & 96.15 & 12.35 & 0.00 & 86.11 \\
\smell{MissingForkGuardPR} & 70.73 & 0.00 & 11.59 & 0.00 & 46.88 \\
\smell{MissingForkGuardSchedule} & 92.47 & 100.00 & 0.00 & 0.00 & 95.83 \\
\smell{MissingJobTimeout} & 73.44 & 71.76 & 75.41 & 64.81 & 73.85 \\
\smell{MissingPathFilters} & 93.33 & 82.35 & 14.29 & 49.06 & 96.67 \\
\smell{MissingPermExternalActions} & 83.76 & 97.96 & 38.71 & 12.99 & 97.03 \\
\smell{MissingTokenPermissions} & 86.96 & 91.59 & 2.86 & 58.82 & 85.47 \\
\smell{MissingUploadCondition} & 50.98 & 0.00 & 92.96 & 0.00 & 28.57 \\
\smell{MultiCommandStep} & 43.75 & 52.94 & 68.42 & 7.69 & 68.42 \\
\smell{MutableRunnerVersion} & 100.00 & 100.00 & 94.34 & 88.17 & 100.00 \\
\smell{SingleLangVersionCI} & 11.54 & 0.00 & 7.02 & 11.54 & 5.00 \\
\smell{SingleOSExecution} & 68.29 & 78.65 & 78.85 & 3.92 & 33.33 \\
\smell{SuboptimalCacheUsage} & 31.03 & 15.09 & 18.52 & 0.00 & 95.74 \\
\smell{UnpinnedActionVersion} & 94.59 & 84.93 & 81.40 & 9.76 & 94.59 \\
\smell{UnpinnedDependencies} & 94.59 & 95.89 & 79.37 & 23.26 & 98.41 \\
\smell{UnsafeArtifactUploadFork} & 98.46 & 0.00 & 95.24 & 0.00 & 84.21 \\
\smell{UnsafeDeployOnFork} & 76.54 & 52.94 & 34.92 & 0.00 & 90.11 \\
\bottomrule
\end{tabular}
\end{table*}

\textbf{Finding 3: Effectiveness varies by smell and by the evidence each criterion requires; no configuration dominates.}
For \smell{MissingConcurrencyPR}, \jev{} reaches 100.00\% F1 and \gptdec{} reaches 96.15\%, compared with 12.35\% for \tevone{} and zero for \gemmafour{}. The ordering changes for \smell{MissingUploadCondition}: \tevone{} achieves 92.96\%, versus 50.98\% for \jev{} and zero for \gptdec{} and \gemmafour{}. All five configurations remain below 12\% F1 on \smell{SingleLangVersionCI}, although \luna{} is evaluated on a subset of pairs. Excluding the semi-synthetic workflows changes the three decision models' overall micro-F1 by less than one point (79.76\%, 71.30\%, 51.42\%), but leaves only five natural positives for \smell{UnsafeDeployOnFork}.

The criteria require different evidence, which helps interpret these contrasts. A mutable runner label is explicit in a single field, whereas \smell{SingleLangVersionCI} depends on compatibility requirements that may be absent from the YAML. \smell{SuboptimalCacheUsage} requires relating a caching step to setup-action capabilities: \luna{} reaches 95.74\% F1 on its classified pairs, while no decision model exceeds 31.03\%. Conversely, the timeout exception illustrated below defeats all five configurations. These cases identify differences in evidence and interpretation; they do not isolate a causal effect of reasoning ability or model architecture.

Listing~\ref{lst:observed-workflow} illustrates model disagreements on \texttt{0xpolygonmiden@miden-vm@changelog.yml}. The workflow invokes a reusable workflow for pull-request events and explicitly restricts token permissions to \texttt{contents: read}. The benchmark labels \smell{MissingConcurrencyPR} as present and \smell{MissingTokenPermissions} and \smell{MissingJobTimeout} as absent. 

\begin{lstlisting}[
    style=gha-motivation,
    morekeywords={
        name,on,jobs,steps,uses,run,if,id,with,
        runs-on,permissions,env,concurrency,
        group,cancel-in-progress,timeout-minutes,
        needs,strategy,matrix,outputs,secrets,environment,
        contents,branches,types
    },
    upquote=true,
    breaklines=true,
    breakatwhitespace=false,
    breakautoindent=true,
    breakindent=0pt,
    postbreak=\mbox{\textcolor{gray}{$\hookrightarrow$}\space},
    columns=fullflexible,
    linewidth=\linewidth,
    captionpos=b,
    caption={Workflow excerpt illustrating model disagreements (reusable-workflow commit hash abbreviated).},
    label={lst:observed-workflow}
]
name: changelog

permissions:
  contents: read

on:
  pull_request:
    branches:
      - main
      - next
    types: [opened, reopened, synchronize,
            labeled, unlabeled]

jobs:
  changelog-reusable:
    uses: 0xMiden/.github/.github/workflows/changelog.yml@d34ae3c
\end{lstlisting}

Table~\ref{tab:case-study} shows that \jev{}, \gptdec{}, and \luna{} match the annotated \smell{MissingConcurrencyPR} label. However, this label depends on context outside the supplied YAML: the caller alone does not establish whether the referenced workflow implements concurrency controls. Both generative configurations incorrectly flag \smell{MissingTokenPermissions}; \luna{} even acknowledges the restricted permissions in its explanation while returning \texttt{YES}. All five configurations incorrectly flag \smell{MissingJobTimeout}. Under our criterion, the caller job is excluded because jobs invoking reusable workflows through \texttt{jobs.<id>.uses} do not support \texttt{timeout-minutes}~\cite{github2026reusable}. The example illustrates both inconsistent verdicts and failures to apply criterion-specific exceptions.

\begin{table}[t]
\centering
\caption{Classifications for Listing~\ref{lst:observed-workflow}. Ref.: reference label (Y: smell present; N: absent). \cmark: prediction agrees with the reference; \xmark: prediction disagrees.}
\label{tab:case-study}
\scriptsize
\setlength{\tabcolsep}{4pt}
\renewcommand{\arraystretch}{0.9}
\begin{tabular}{@{}lcccccc@{}}
\toprule
\textbf{Smell} & \textbf{Ref.} & \textbf{Jev} & \textbf{GPT} & \textbf{Tev1} & \textbf{Gemma} & \textbf{Luna} \\
\midrule
\smell{MissingTokenPermissions} & N & \cmark & \cmark & \cmark & \xmark & \xmark \\
\smell{MissingJobTimeout}       & N & \xmark & \xmark & \xmark & \xmark & \xmark \\
\smell{MissingConcurrencyPR}    & Y & \cmark & \cmark & \xmark & \xmark & \cmark \\
\bottomrule
\end{tabular}
\end{table}

The scope grouping in Table~\ref{tab:catalog} distinguishes ten localized and nine contextual criteria, comprising 834 and 876 pairs. Localized versus contextual micro-F1 is 80.34\% versus 78.98\% for \jev{}, 73.78\% versus 67.34\% for \gptdec{}, 66.84\% versus 30.72\% for \tevone{}, and 40.06\% versus 9.52\% for \gemmafour{}. The larger gaps for the local configurations are consistent with the difficulties on relational and applicability-dependent criteria illustrated above. However, the groups contain different smells and label distributions, so these are descriptive contrasts rather than controlled tests of contextual complexity. Localized checks can also require exceptions, as the timeout example demonstrates.

\textbf{Finding 4: Score discrimination and binary classification reveal distinct failure modes.}
Per-smell AUROC separates two failure modes at the default threshold. \gptdec{}'s four zero-F1 smells have AUROC 0.83--1.00 (1.00 on \smell{UnsafeArtifactUploadFork}), indicating useful score discrimination that the 0.5 cutoff does not translate into positive detections. By contrast, \tevone{}'s AUROC falls below 0.5 on four smells, down to 0.146 on \smell{MissingTokenPermissions}, while the same definitions yield 0.93--1.00 for \jev{}: positive pairs tend to receive lower scores than negative ones. 

\begin{figure}[ht!]
\centering
\begin{tikzpicture}[font=\scriptsize]
\definecolor{overlapInk}{RGB}{76,126,164}
\definecolor{overlapGray}{RGB}{215,220,225}
\def\colw{0.405}   
\def\xzero{2.05}   
\def\rowgap{0.34}  

\draw[gray!50] ({\xzero-0.22},0) -- ({\xzero+14*\colw+0.22},0);
\foreach \name [count=\row from 1] in {\jev{},\gptdec{},\tevone{},\gemmafour{}}
  \node[anchor=east] at ({\xzero-0.25},{-0.30-\rowgap*(\row-1)}) {\name};

\foreach \col/\fj/\fg/\ft/\fm/\paircount in {
0/1/1/0/0/183, 1/0/0/0/0/167, 2/1/1/1/0/156, 3/1/1/1/1/88,
4/1/0/0/0/77,  5/1/0/1/0/58,  6/1/1/0/1/54,  7/0/0/1/0/39,
8/0/1/0/0/14,  9/0/1/1/0/9,   10/0/0/0/1/4,  11/1/0/0/1/3,
12/0/0/1/1/1,  13/0/1/0/1/1,  14/1/0/1/1/1
}{
  \pgfmathsetmacro{\x}{\xzero+\colw*\col}
  \pgfmathsetmacro{\barheight}{0.0085*\paircount}
  \fill[overlapInk] (\x-0.13,0) rectangle (\x+0.13,\barheight);
  \node[anchor=south,font=\tiny\bfseries] at (\x,\barheight+0.02) {\paircount};

  \pgfmathtruncatemacro{\first}{\fj==1 ? 1 : (\fg==1 ? 2 : (\ft==1 ? 3 : (\fm==1 ? 4 : 0)))}
  \pgfmathtruncatemacro{\last}{\fm==1 ? 4 : (\ft==1 ? 3 : (\fg==1 ? 2 : (\fj==1 ? 1 : 0)))}
  \ifnum\first>0
    \ifnum\last>\first
      \draw[overlapInk,line width=1pt]
        (\x,{-0.30-\rowgap*(\first-1)}) -- (\x,{-0.30-\rowgap*(\last-1)});
    \fi
  \else
    \node[anchor=north,font=\tiny] at (\x,{-0.30-\rowgap*3-0.12}) {None};
  \fi
  \foreach \active [count=\row from 1] in {\fj,\fg,\ft,\fm} {
    \pgfmathsetmacro{\y}{-0.30-\rowgap*(\row-1)}
    \ifnum\active=1
      \fill[overlapInk] (\x,\y) circle (2.1pt);
    \else
      \fill[overlapGray] (\x,\y) circle (1.5pt);
    \fi
  }
}
\end{tikzpicture}
\caption{Exclusive detection patterns across 855 positive pairs. Blue circles indicate detection and gray circles indicate misses; ``None'' denotes pairs missed by all four.}
\label{fig:upset-positive}
\end{figure}
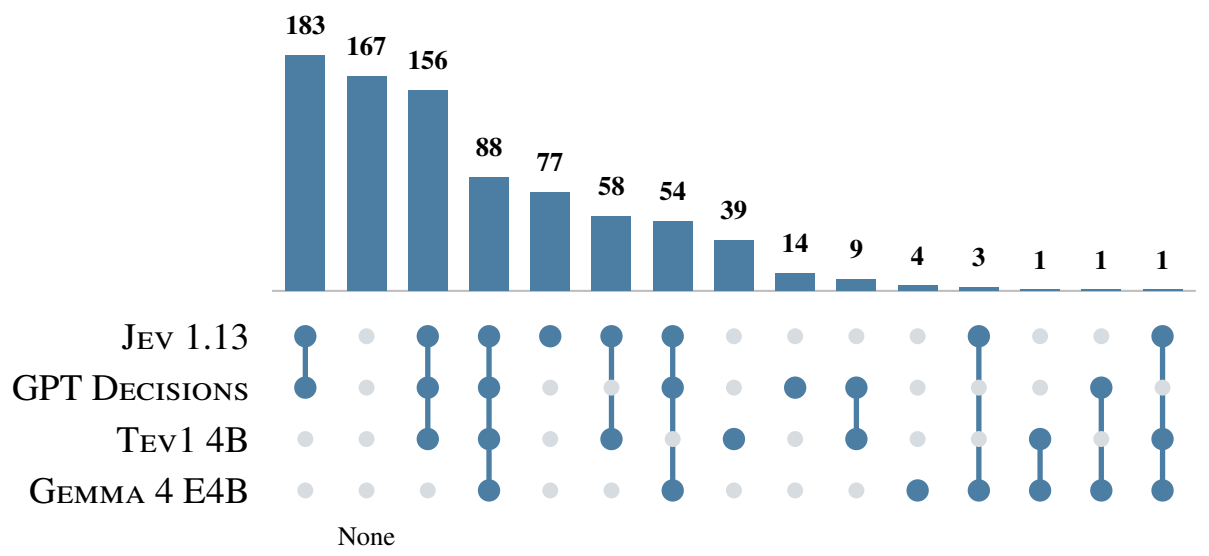

\textbf{Finding 5: Complementary detections do not necessarily improve combined effectiveness.}
Figure~\ref{fig:upset-positive} examines complementary detections among the four full-coverage configurations. Each bar counts positive pairs detected by exactly the models marked with blue circles below it; gray circles indicate models that missed those pairs. For example, the bar labeled 183 represents pairs detected by \jev{} and \gptdec{} but missed by \tevone{} and \gemmafour{}. The column labeled ``None'' represents 167 pairs missed by all four models. Overall, at least one model detects 688 of the 855 positive pairs, and all four detect 88. The zero-count combination is omitted.

Exclusive detections comprise 77 positive pairs for \jev{}, 14 for \gptdec{}, 39 for \tevone{}, and four for \gemmafour{}. Including negative pairs, the corresponding counts of classifications correct for exactly one model are 77, 23, 46, and nine. Thus, some exclusive successes reflect correctly rejecting an absent smell. For each annotated pair, we compare all four binary predictions with its reference label. Counting pairs with four correct predictions yields 682; counting pairs with four incorrect predictions yields 195. 

The latter comprise the 167 positive pairs missed by all models and 28 negative pairs flagged by all models.
An OR rule, which predicts presence whenever any model does, yields 688 true positives, 261 false positives, and 167 false negatives, producing 76.27\% micro-F1, below \jev{} alone (79.69\%). An AND rule, which requires all four models to predict presence, yields 88 true positives, 28 false positives, and 767 false negatives, producing 18.13\% micro-F1. These exploratory combinations illustrate the trade-off between recovering missed smells and introducing false alarms.

\begin{rqanswer}{2}
Effectiveness varies substantially across smells: no configuration dominates, local models struggle on contextual criteria, and combining predictions adds false alarms faster than detections.
\end{rqanswer}

\subsection{Decision Thresholds and Abstentions (\textbf{RQ$_3$})}
\label{sec:rq3-results}

We examine how threshold selection and abstention affect detection effectiveness. Table~\ref{tab:threshold} compares five thresholds for the three decision models on the same 1,710 annotated pairs. The generative configurations return categorical judgments rather than comparable numerical scores and are therefore excluded from this sweep. Threshold 0.5 is the primary setting; the alternatives are explored retrospectively.

\begin{table}[ht!]
\centering
\caption{Exploratory micro-F1 (\%) across decision thresholds. Bold values mark the best inspected threshold per model.}
\label{tab:threshold}
\footnotesize
\setlength{\tabcolsep}{7pt}
\begin{tabular}{@{}rrrr@{}}
\toprule
\textbf{Threshold} &
\textbf{\jev{}} &
\textbf{\gptdec{}} &
\textbf{\tevone{}} \\
\midrule
0.2 & 84.92 & \textbf{77.23} & \textbf{74.81} \\
0.3 & \textbf{85.54} & 75.23 & 71.49 \\
0.4 & 83.66 & 72.80 & 61.67 \\
0.5 & 79.69 & 70.68 & 50.61 \\
0.6 & 71.00 & 69.32 & 34.97 \\
\bottomrule
\end{tabular}
\end{table}

\textbf{Finding 6: Lower thresholds improve effectiveness, and repository-grouped selection yields within-benchmark gains that narrow the gaps between models.} Relative to the default threshold of 0.5, the lower thresholds inspected improve micro-F1 for all three models. For \jev{}, reducing the threshold from 0.5 to 0.3 increases F1 from 79.69\% to 85.54\%: recall rises from 72.51\% to 87.84\%, while precision decreases from 88.45\% to 83.35\%. At threshold 0.2, \gptdec{} reaches 77.23\% F1 and \tevone{} reaches 74.81\%, compared with 70.68\% and 50.61\% at 0.5. These results demonstrate sensitivity to the cutoff, but selecting and evaluating thresholds on the same data can overestimate their benefit.

To separate threshold selection from evaluation, we use five-fold cross-validation grouped by repository, keeping pairs from the same repository in one fold. For each held-out fold, we select the threshold maximizing micro-F1 on the other four folds, searching from 0.05 to 0.95 in increments of 0.01 and breaking ties by proximity to 0.5. We then pool predictions from the held-out folds. Only the F1 columns in Table~\ref{tab:decision-discrimination} use cross-validated threshold selection. The table also reports AUROC and average precision (AP), which summarize how well scores rank positive pairs above negative pairs across thresholds; higher values indicate better discrimination.

\begin{table}[ht!]
\centering
\caption{Score discrimination, calibration error, and repository-grouped micro-F1 (\%). AP: average precision; F1$_s$: per-smell thresholds.}
\label{tab:decision-discrimination}
\footnotesize
\setlength{\tabcolsep}{3pt}
\begin{tabular}{@{}lrrrrrr@{}}
\toprule
\textbf{Model} & \textbf{AUROC} & \textbf{AP} & \textbf{Brier} & \textbf{ECE} & \textbf{CV F1} & \textbf{CV F1$_s$} \\
\midrule
\jev{} & 0.910 & 0.867 & 0.134 & 0.118 & 85.43 & 90.38 \\
\gptdec{} & 0.877 & 0.851 & 0.201 & 0.203 & 81.68 & 84.58 \\
\tevone{} & 0.716 & 0.656 & 0.230 & 0.154 & 74.46 & 77.47 \\
\bottomrule
\end{tabular}
\end{table}

Cross-validated threshold selection increases micro-F1 by 5.74 percentage points for \jev{}, 11.00 for \gptdec{}, and 23.85 for \tevone{}. The \jev{}--\gptdec{} gap consequently narrows from 9.01 to 3.75 points, while the \jev{}--\tevone{} gap narrows from 29.08 to 10.97 points. Thus, part of the default-threshold performance gap reflects the decision policy. Fold assignments are fixed and archived; across 20 assignments, \jev{}'s micro-F1 ranges from 84.71\% to 85.84\%. Assigning repositories that share identical YAML files to the same fold (moving 22 repositories) leaves the cross-validated micro-F1 of all three decision models unchanged. The \jev{}--\gptdec{} gap has a 95\% CI of [1.28, 6.31] points. For \tevone{}, threshold selection favors recall (94.62\%) at the expense of precision (61.38\%), illustrating the trade-off behind its F1 improvement. Per-smell thresholds further raise micro-F1 (Table~\ref{tab:decision-discrimination}), so a global cutoff conceals differences in score scale across smells.

Selected thresholds range from 0.23--0.26 for \jev{}, 0.05--0.06 for \gptdec{}, and 0.20--0.23 for \tevone{}. Extending the grid down to 0.01 selects thresholds of 0.02--0.03 for \gptdec{} and raises its cross-validated micro-F1 to 82.22\%, so the reported value is slightly conservative. Low F1-optimal thresholds alone do not establish poor probability calibration, since the objective and class distribution also affect the cutoff~\cite{lipton2014thresholding,guo2017calibration}, so we also report squared probability error (Brier) and discrepancies between mean scores and positive rates across bins (ECE); lower values are better (Table~\ref{tab:decision-discrimination}). The observed ECE values caution against treating scores as calibrated probabilities on this benchmark. These estimates assess threshold selection within the existing benchmark; RQ7 examines their transfer.

\textbf{Finding 7: Abstention changes the interpretation of detection effectiveness by reducing classification coverage.} For \luna{}, the relevant policy is abstention: 1,463 pairs receive binary judgments, 245 receive \texttt{UNCERTAIN}, and two requests fail. Its 80.53\% micro-F1 therefore describes the 85.56\% of pairs classified, leaving effectiveness on the remaining pairs unknown. If all 247 non-decisions were resolved incorrectly or correctly, full-coverage F1 would be 68.62\% or 83.26\%, respectively. These bounds illustrate the uncertainty introduced by incomplete coverage; they are not observed classification results. Counting abstentions as negative or positive predictions yields full-coverage micro-F1 of 74.46\% or 77.16\%, both below \jev{}’s 79.69

To examine effectiveness at the same coverage, we retrospectively defer the 247 \jev{} predictions whose scores are closest to 0.5. On the remaining 1,463 pairs, \jev{} achieves 81.92\% micro-F1, compared with 79.69\% at full coverage. Although this exceeds \luna{}'s conditional F1 of 80.53\% (Table~\ref{tab:performance}), the retained subsets differ, so equal coverage does not establish superiority on identical cases. This exploratory policy illustrates the trade-off between classification coverage and effectiveness; its deferral budget was not selected on independent data.

\begin{rqanswer}{3}
Decision policies substantially affect measured effectiveness: cross-validated thresholds improve micro-F1, while abstention reduces coverage and changes the interpretation of reported F1.
\end{rqanswer}

\subsection{Cost, Token Usage, and Latency (\textbf{RQ$_4$})}
\label{sec:rq4-results}

We assess operational efficiency through API charges, token usage, and request latency. The decision configurations and \gemmafour{} batch smell checks by workflow, whereas \luna{} evaluates one annotated workflow--smell pair per request. Per-request measurements therefore represent different amounts of work. For \luna{}, token usage and charges cover 1,708 completed responses.

\noindent\textbf{Finding 8: \jev{} has the lowest hosted API charges and median latency, although batching and workload coverage limit cost comparisons.} As Table~\ref{tab:cost} shows, \jev{} incurs the lowest aggregate API charge among hosted configurations (US\$0.224). Although \luna{} costs less than \gptdec{} in this run (US\$0.255 versus US\$0.763), it evaluates only annotated pairs, whereas \gptdec{} evaluates all catalog criteria for each workflow. These totals reflect different workloads; provider-reported token counts are not normalized computational costs. Local models incur no metered API charges, but electricity, hardware acquisition and depreciation, and administration costs were not measured. \jev{}'s output-token count reflects provider-reported usage metadata rather than generated natural-language explanations. The \gptdec{} charges include 97,208 cache-write tokens at US\$0.125 per million. For \tevone{}, reported input usage averages 836 tokens per question; the largest workflow-level average is 3,158.

\begin{table}[ht!]
\centering
\caption{Observed token usage and estimated API charges (US\$) under the evaluated workloads. Local operating costs are excluded.
}
\label{tab:cost}
\footnotesize
\setlength{\tabcolsep}{5pt}
\resizebox{\linewidth}{!}{

\begin{tabular}{@{}lrrrrrrr@{}}
\toprule
\textbf{Model} &
\textbf{Requests} &
\textbf{Input tokens} &
\textbf{Output tokens} &
\textbf{Total US\$} &
\textbf{Min US\$} &
\textbf{Median US\$} &
\textbf{Max US\$} \\
\midrule
\jev{} & 903 & 5,334,268 & 400,932 & \textbf{0.224} & 0.000229 & 0.000244 & 0.000343 \\
\gptdec{} & 903 & 7,605,732 & 0 & 0.763 & 0.000801 & 0.000833 & 0.001115 \\
\tevone{} & 903 & 16,610,044 & 0 & 0 API & -- & -- & -- \\
\gemmafour{} & 903 & 1,860,883 & 139,369 & 0 API & -- & -- & -- \\
\luna{} & 1,710 & 1,324,625 & 244,692 & 0.255 & 0.000068 & 0.000135 & 0.000476 \\
\bottomrule
\end{tabular}
}
\end{table}

The hosted decision configurations achieve median latencies below one second per workflow (Table~\ref{tab:latency}). Local \tevone{} takes approximately eight times as long as \jev{} at the median (5.430 versus 0.675 seconds), while \gemmafour{} takes 1.547 seconds. The 17 \tevone{} requests below one second repeat an earlier YAML. These measurements characterize the evaluated deployments, including their hardware and serving environments.

\begin{table}[ht!]
\centering
\caption{Request latency (seconds). P95 is the 95th percentile; totals sum recorded request durations, including failed calls.}
\label{tab:latency}
\scriptsize
\setlength{\tabcolsep}{4pt}
\begin{tabular}{@{}lrrrrr@{}}
\toprule
\textbf{Model} &
\textbf{Total} &
\textbf{Min} &
\textbf{Median} &
\textbf{P95} &
\textbf{Max} \\
\midrule
\jev{} & 623.3 & 0.581 & \textbf{0.675} & 0.783 & 4.108 \\
\gptdec{} & 949.4 & 0.667 & 0.978 & 1.441 & 6.273 \\
\tevone{} & 4,968.7 & 0.042 & 5.430 & 6.900 & 9.527 \\
\gemmafour{} & 1,553.4 & 0.784 & 1.547 & 3.106 & 4.539 \\
\luna{} & 6,260.2 & 1.627 & 3.083 & 7.596 & 17.534 \\
\bottomrule
\end{tabular}
\end{table}

Batching also affects the cost comparison. Each hosted decision request evaluates 22 catalog questions, including the 19 scored smells, whereas \luna{} evaluates only the annotated pairs, averaging 1.89 smells per workflow, and also generates explanations. At its observed average charge per pair, evaluating all 19 smells across 903 workflows would cost approximately US\$2.56, assuming unchanged token usage and pricing per pair. This extrapolation illustrates the cost of extending coverage; a complete-catalog \luna{} run was not measured.

\begin{rqanswer}{4}
In our setup, \jev{} had the lowest API charges and the shortest accumulated request latency among the hosted configurations under the evaluated workloads, although batching differences limit direct cost comparisons and local operating costs remain unmeasured.
\end{rqanswer}

\subsection{Stability Across Repeated Inference (\textbf{RQ$_5$}) }
\label{sec:rq5-results}

We assess whether repeated inference produces consistent labels and whether majority voting improves effectiveness. We ran each decision configuration three times on the same 1,710 annotated pairs, keeping workflows, smell definitions, and the threshold of 0.5 unchanged.
We measure complete label agreement across runs, variation in run-level accuracy, and micro-F1 after majority voting. This analysis complements evidence that nominally deterministic language-model settings can vary~\cite{atil2025nondeterminismdeterministicllmsettings}.

\textbf{Finding 9: Decision models are highly repeatable, but repetition and majority voting add cost without improving effectiveness.} Table~\ref{tab:decision-stability-results} shows that \jev{} produces identical labels across all three runs for 1,661 pairs (97.13\%). Each of the 49 remaining pairs has at least one score between 0.4 and 0.6, placing the observed disagreements near the decision threshold. Score variation is small but pervasive for \jev{}: scores differ across runs on 76.5\% of pairs, with a median range of 0.01 and a 95th percentile of 0.07; only 28 pairs vary by more than 0.1 (maximum 0.31). Labels therefore change only when scores lie close to the threshold. Both \gptdec{} and \tevone{} produce identical labels across all runs. \gptdec{} also returns identical numerical scores across all three runs; the second run reports no cached input tokens. For \tevone{}, scores vary on 41 pairs, by at most 0.021, without crossing the threshold. High repeatability therefore coexists with substantially different detection effectiveness.

\begin{table}[ht!]
\centering
\caption{Agreement (\%), accuracy spread (percentage points), and micro-F1 (\%) across three runs.}
\label{tab:decision-stability-results}
\footnotesize
\setlength{\tabcolsep}{4pt}
\begin{tabular}{@{}lrrrr@{}}
\toprule
\textbf{Model} &
\textbf{Agreement} &
\textbf{Accuracy} &
\multicolumn{2}{c}{\textbf{Micro-F1}} \\
\cmidrule(l){4-5}
& \textbf{(\%)} & \textbf{spread} & \textbf{Run 1} & \textbf{Majority} \\
\midrule
\jev{} & 97.13 & 0.23 & 79.69 & 79.64 \\
\gptdec{} & 100.00 & 0.00 & 70.68 & 70.68 \\
\tevone{} & 100.00 & 0.00 & 50.61 & 50.61 \\
\bottomrule
\end{tabular}
\end{table}

Majority voting provides no micro-F1 improvement for any configuration. For \jev{}, it yields 79.64\%, compared with 79.69\% in the first run; the other models retain their original predictions. Although repetition produces at least one correct prediction for 24 pairs initially misclassified by \jev{}, identifying those correct outputs would require knowledge of the reference labels. These potential corrections do not translate into a net benefit from majority voting.

Repetition also increases resource use. Estimated API charges total US\$0.672 for three \jev{} runs and US\$2.278 for three \gptdec{} runs, compared with US\$0.224 and US\$0.763 for their first runs. Charges across the three runs total US\$2.278 rather than three times US\$0.763 because the third run incurred no cache-write charges. The three local \tevone{} runs accumulate approximately 4.16 hours of request durations. Thus, the evaluated strategy roughly triples API charges or inference time without improving majority-vote F1.

Disagreement nevertheless identifies a subset with a higher observed error rate. Of the 49 unstable \jev{} pairs, 24 were incorrect in the first run: 49.0\%, compared with 17.6\% among stable pairs. However, these cases account for only 7.6\% of its 316 first-run errors. Disagreement may therefore help prioritize review, but detects only a small fraction of errors in this experiment.

\begin{rqanswer}{5}
Decision models are highly repeatable (97--100\% label agreement), but repetition and majority voting add cost without improving effectiveness, and most errors are stable across runs.
\end{rqanswer}

\subsection{Robustness under Metamorphic Testing (\textbf{RQ$_6$})}
\label{sec:metamorphic-robustness}

We use metamorphic testing to assess whether predictions remain consistent after edits that should preserve the expected classification. Each workflow receives one of five transformations: adding comments, changing formatting, reordering mapping keys and jobs, consistently renaming jobs, or changing equivalent condition delimiters. These transformations preserve the presence or absence of the assessed smells, leaving their reference labels unchanged. We evaluate each decision model once on the transformed benchmark and use the three original-input runs from \textbf{RQ$_5$} as a baseline for inference variability. We count changed labels among the 1,710 pairs; the repetition baseline is the mean number of label changes between original runs. For \jev{}, we average metamorphic flips over comparisons with its three original runs.

\textbf{Finding 10: Label-preserving edits change aggregate micro-F1 by less than 0.6 percentage points but flip individual predictions, even for perfectly repeatable models.} Table~\ref{tab:metamorphic-results} shows little change in aggregate effectiveness: micro-F1 differs by less than 0.6 percentage points from each model's first original run. All 90\% confidence intervals, estimated using 10,000 paired repository-cluster bootstrap resamples, lie within the $\pm2$-point equivalence margin. However, aggregate stability within this exploratory margin can conceal individual prediction changes when corrections and newly introduced errors offset one another.

\begin{table}[ht!]
\centering
\caption{Original and transformed micro-F1 (\%) and label flips. Differences and 90\% bootstrap CIs are in percentage points; Rep.\ denotes the repetition baseline. For \jev{}, flips are averaged over comparisons with its three original runs.}
\label{tab:metamorphic-results}
\footnotesize
\setlength{\tabcolsep}{2.5pt}
\begin{tabular}{@{}lrrrcrr@{}}
\toprule
\textbf{Model} & \textbf{Orig.} & \textbf{Transf.} & \textbf{$\Delta$} &
\textbf{90\% CI} & \textbf{Flips} & \textbf{Rep.} \\
\midrule
\jev{}    & 79.69 & 79.77 & $+0.08$ & [$-0.69$, $0.81$] & 47 & 32.7 \\
\gptdec{} & 70.68 & 71.26 & $+0.58$ & [$-0.39$, $1.58$] & 54 & 0 \\
\tevone{} & 50.61 & 50.14 & $-0.47$ & [$-1.51$, $0.54$] & 61 & 0 \\
\bottomrule
\end{tabular}
\end{table}

For \gptdec{}, 54 labels change (3.16\%): 30 incorrect predictions become correct, while 24 correct predictions become incorrect. This contrasts with identical scores across its three original runs. Changes are not confined to borderline predictions: 20 of the 54 flipped pairs have original scores outside [0.3, 0.7], and the median absolute score change among flipped pairs is 0.38. For example, adding a neutral comment changes an \smell{UnpinnedActionVersion} score from 0.03 to 0.76 without modifying any action reference. All \gptdec{} runs used the generally available API: the original runs were executed on 28 October 2026 and the transformed run on 29 October, about 26 hours after the last original run. Because original and transformed runs were not executed on the same day, a provider-side update cannot be excluded as an alternative explanation.

For \tevone{}, 61 labels change (3.57\%), comprising 30 corrections and 31 newly incorrect predictions, despite complete label agreement across its original runs. All flipped pairs have original scores within [0.3, 0.7], with a median absolute score change of 0.056. Its observed sensitivity therefore mainly reflects small score changes crossing the decision threshold.
For \jev{}, original runs differ on 32.7 labels on average (1.91\%), compared with 47.0 labels (2.75\%) between original and transformed runs, an increase of 0.84 percentage points. Among the 1,661 pairs labeled consistently across all three original runs, 23 change after transformation. All 23 have original scores within [0.3, 0.7]. These comparisons suggest additional sensitivity to the edits, although one transformed run cannot fully separate it from inference variability.

\textbf{Finding 11: Prediction instability varies across transformations and decision thresholds.} Mapping-key and job reordering (T3) produces the most observed flips for \tevone{} (25 of 61) and \jev{} (on average 16.3 of 47.0 per run). For \gptdec{}, flips are distributed more evenly, with 10--13 in each transformation group. Because the transformations are applied to different workflow groups, these counts do not establish which transformation is intrinsically more disruptive. Comparing transformation types or sensitivity across local and contextual smells requires further analysis. The number of label flips also depends on the operating point. At thresholds selected on the full benchmark (0.26, 0.06, and 0.23), flips are 40.3 for \jev{} (repetition baseline 30.0), 96 for \gptdec{} (0), and 46.7 for \tevone{} (0.7). \gptdec{}'s rate thus nearly doubles to 5.61\% at this low cutoff.

\begin{rqanswer}{6}
Repeatability does not imply robustness: label-preserving edits change aggregate micro-F1 by less than 0.6 percentage points but flip 2.8--3.6\% of individual predictions, including for models with perfect repeatability.
\end{rqanswer}

\subsection{Comparison with a Static Analyzer (\textbf{RQ$_7$})}
\label{sec:external-validation}

We compare decision models with Khatami et al.'s static analyzer~\cite{khatami2024catching} using manual assessments from its original evaluation. The historically reconstructed set contains 46 workflows and 874 workflow--smell pairs, including 275 positive labels. As described in Section~\ref{sec:external-design}, reconstruction retained revisions whose detector outputs matched the recorded results; this does not guarantee recovery of the original commits and may favor the static analyzer. Decision models use threshold 0.5. \tevone{} could not evaluate one workflow (\texttt{prisma@prisma@test-template.yml}, 1,261 lines): with the 22 catalog questions, the request exceeded the 64~KB limit that the local Ollama server imposes on prompt text and response schema. Comparisons involving \tevone{} therefore use the remaining 855 pairs, on which the static analyzer reaches 88.2\% micro-F1.

\textbf{Finding 12: The static analyzer outperforms the decision models on the reconstructed set, and thresholds selected on the main benchmark do not close the gap.} Table~\ref{tab:external-overall} shows higher micro-F1 for the static analyzer than for each decision model, with all paired confidence intervals excluding zero. Its advantage is especially pronounced in recall: it identifies 97.8\% of positive pairs, compared with 81.8\% for \jev{} and 76.4\% for \gptdec{}. The difference between these two models is smaller than on the main benchmark: 1.7 percentage points (95\% CI: [$-1.4$, $5.1$]), providing insufficient evidence of a difference on this set.

\begin{table}[ht!]
\centering
\caption{Effectiveness on reconstructed workflows (\%). $\Delta$ is model minus static micro-F1 (95\% repository-cluster bootstrap CI).$^\dagger$ Results use 855 pairs.}
\label{tab:external-overall}
\footnotesize
\setlength{\tabcolsep}{2.5pt}
\begin{tabular}{@{}lrrrrr@{}}
\toprule
\textbf{Detector} & \textbf{P} & \textbf{R} & \textbf{Micro-F1} & \textbf{Macro-F1} & \textbf{$\Delta$ micro-F1} \\
\midrule
Static~\cite{khatami2024catching} & 79.6 & 97.8 & 87.8 & 82.6 & --- \\
\jev{} & 71.7 & 81.8 & 76.4 & 64.7 & $-11.4$ [$-16.6$, $-6.7$] \\
\gptdec{} & 73.2 & 76.4 & 74.7 & 54.4 & $-13.0$ [$-19.8$, $-6.8$] \\
\tevone{}$^\dagger$ & 62.8 & 66.8 & 64.8 & 43.9 & $-23.4$ [$-30.7$, $-17.5$] \\
\bottomrule
\end{tabular}
\end{table}

The comparison varies with smell scope (Table~\ref{tab:external-smell}). The static analyzer achieves 97.3\% micro-F1 on localized criteria, whereas all detectors perform worse on contextual criteria. For the latter, the static analyzer reaches 69.8\%, compared with 62.0\% for \jev{}; their difference is $-7.8$ points (95\% CI: [$-17.5$, $0.9$]), which establishes neither a difference nor equivalence. Individual model advantages also occur: \jev{} achieves higher F1 on \smell{MissingConcurrencyPR}, and \gptdec{} on \smell{SingleOSExecution}. These results concern small numbers of positive instances and should be interpreted accordingly.

\begin{table}[ht!]
\centering
\caption{Per-smell F1 (\%) on reconstructed workflows. Pos.\ counts positive reference labels across 46 workflows; group rows report pooled micro-F1. \tevone{} excludes one failed workflow, so its support differs.}
\label{tab:external-smell}
\footnotesize
\setlength{\tabcolsep}{3pt}
\resizebox{\columnwidth}{!}{%
\begin{tabular}{@{}lrrrrr@{}}
\toprule
\textbf{Smell} & \textbf{Pos.} & \textbf{Static} & \textbf{\jev{}} & \textbf{\gptdec{}} & \textbf{\tevone{}} \\
\midrule
\multicolumn{6}{@{}l}{\emph{Localized}} \\
\smell{MutableRunnerVersion}       & 43 & 100 & 99 & 99 & 98 \\
\smell{MissingJobTimeout}          & 42 & 100 & 99 & 99 & 96 \\
\smell{UnpinnedActionVersion}      & 37 & 99 & 95 & 96 & 95 \\
\smell{MissingPermExternalActions} & 24 & 100 & 75 & 98 & 61 \\
\smell{MultiCommandStep}           & 24 & 100 & 59 & 67 & 69 \\
\smell{MissingTokenPermissions}    & 8 & 100 & 46 & 52 & 0 \\
\smell{UnsafeArtifactUploadFork}   & 7 & 67 & 62 & 25 & 57 \\
\smell{SuboptimalCacheUsage}       & 5 & 100 & 67 & 57 & 55 \\
\smell{MissingUploadCondition}     & 5 & 83 & 75 & 0 & 40 \\
\smell{UnpinnedDependencies}       & 3 & 60 & 33 & 32 & 55 \\
\emph{All localized} & 198 & \textbf{97.3} & 82.3 & 85.1 & 79.6 \\
\midrule
\multicolumn{6}{@{}l}{\emph{Contextual}} \\
\smell{MissingConcurrencyGlobal}   & 16 & 100 & 60 & 11 & 0 \\
\smell{MissingConcurrencyBranch}   & 11 & 96 & 85 & 54 & 64 \\
\smell{MissingConcurrencyPR}       & 11 & 80 & 88 & 71 & 0 \\
\smell{MissingPathFilters}         & 10 & 80 & 57 & 48 & 0 \\
\smell{UnsafeDeployOnFork}         & 9 & 35 & 31 & 20 & 18 \\
\smell{MissingForkGuardPR}         & 7 & 88 & 53 & 0 & 35 \\
\smell{SingleOSExecution}          & 7 & 47 & 47 & 57 & 36 \\
\smell{MissingForkGuardSchedule}   & 3 & 86 & 100 & 100 & 0 \\
\smell{SingleLangVersionCI}        & 3 & 50 & 0 & 50 & 57 \\
\emph{All contextual} & 77 & \textbf{69.8} & 62.0 & 48.8 & 26.8 \\
\bottomrule
\end{tabular}%
}
\end{table}

Table~\ref{tab:external-frozen} evaluates thresholds selected on the main benchmark and frozen for external testing.
Global thresholds selected on all 1,710 pairs (0.26, 0.06, and 0.23) reduce external micro-F1 for every model: recall rises, but precision falls more sharply. Per-smell thresholds recover \jev{}'s default result and improve \gptdec{}, yet remain below static analysis.

\begin{table}[ht!]
\centering
\caption{External micro-F1 (\%) with benchmark-selected thresholds. $\Delta$: model minus static (95\% repository-cluster bootstrap CI);$^\dagger$ 855 pairs.}
\label{tab:external-frozen}
\footnotesize
\setlength{\tabcolsep}{3pt}
\resizebox{\columnwidth}{!}{%
\begin{tabular}{@{}lrrrr@{}}
\toprule
\textbf{Model} & \textbf{Global} & \textbf{$\Delta$} & \textbf{Per smell} & \textbf{$\Delta$} \\
\midrule
\jev{} & 73.02 & $-14.75$ [$-19.9$, $-10.1$] & 76.66 & $-11.10$ [$-14.1$, $-8.6$] \\
\gptdec{} & 73.32 & $-14.44$ [$-19.3$, $-9.5$] & 78.65 & $-9.11$ [$-13.5$, $-5.1$] \\
\tevone{}$^\dagger$ & 58.71 & $-29.47$ [$-33.6$, $-26.2$] & 61.17 & $-27.02$ [$-30.7$, $-24.1$] \\
\bottomrule
\end{tabular}%
}
\end{table}

\textbf{Finding 13: Model-based filtering reveals complementary errors, but post-hoc improvements remain unvalidated.} To examine potential complementarity, we inspect cases where models correct static-analysis errors. At threshold 0.5, \jev{} rejects 49 of the analyzer's 69 false alarms but also 46 of its 269 true positives; \gptdec{} rejects 53 and 61. For \smell{UnsafeDeployOnFork}, only eight of the analyzer's 37 findings are true positives: \jev{} rejects 27 of the 29 false alarms but retains only two of the eight true positives, preventing an F1 advantage. An exploratory filter that requires model agreement only for positive \smell{UnsafeDeployOnFork} and \smell{SingleOSExecution} findings raises overall micro-F1 to 92\% with \jev{} and 92.2\% with \gptdec{}, from 87.8\% for static analysis alone. With \tevone{}, it rises from 88.2\% to 91.4\% on the matched subset. Because the smells and filtering rule were chosen after inspecting the reference labels, these gains illustrate complementary errors rather than validated hybrid performance.

These results favor the static analyzer for the implemented checks on this reconstructed set, while the error patterns identify candidates for independently evaluated filtering. The broader motivation for definition-driven detection is specifying additional checks without dedicated rules. This experiment does not evaluate new definitions, evolving syntax, or implementation effort, so it cannot establish an extensibility advantage.

\begin{rqanswer}{7}
The static analyzer outperforms the decision models on the reconstructed workflows, and benchmark-selected thresholds do not close the gap; models reject some static false alarms, but at a cost in recall.
\end{rqanswer}

\section{Discussion}
\label{sec:discussion}

We discuss the implications of our findings for model selection, decision policies, and workflow analysis tools.

\subsection{Decision Interfaces Versus Generative Judgment}

The results from RQ1--RQ4 suggest that configuration selection should consider the required checks, expected outputs and inference costs. Decision-oriented inference provides low-latency classification in our evaluation, while generative inference can additionally provide explanations when requested. However, effectiveness varies across smells (RQ2), and the reusable-workflow example shows that an explanation can contradict its verdict (Section~\ref{sec:rq2-results}). Explanations therefore require review rather than being treated as independent confirmation.

For context, the study that produced our benchmark evaluated the same target-conditioned detection task with agentic generative pipelines on a balanced 570-pair subset of these pairs~\cite{uchoa2026gha}. With a dedicated detection agent, its four LLMs reached 86.53--88.52\% F1 on that subset, compared with \jev{}'s 85.43\% on the full benchmark with cross-validated thresholds, with median latencies of 9.7--21.9 seconds and median costs of US\$0.0006--0.015 per workflow--smell pair, figures that also include refactoring for predicted positives. By contrast, one \jev{} request evaluates all 22 questions for a workflow in a median of 0.675 seconds and US\$0.00024. Because the evaluated subsets and model inputs differ, these figures suggest an effectiveness--cost trade-off rather than a controlled comparison.

This trade-off, together with the per-smell differences, threshold analysis, and latency measurements (RQ2--RQ4), motivates testing a strategy that uses \jev{} for initial screening and selectively invokes a generative model when an explanation is needed or a score falls near the decision threshold. Per-smell results could also inform which checks benefit from additional analysis. However, proximity to the threshold does not identify every error, and a second model may reinforce an incorrect judgment. We have not evaluated this strategy; its effectiveness, review effort, and additional costs require independent assessment.

Our experiments do not isolate the effect of the inference interface on detection effectiveness. Even \gptdec{} and \luna{}, which share a model identifier, use different endpoints, prompts, batching arrangements, and response policies. The per-smell differences in RQ2 motivate testing whether intermediate reasoning helps with relationships across workflow constructs, but do not establish that explanation. A controlled comparison of decision-score readout and generated verdicts from the same open model~\cite{geuter2026inner}, with matched inputs and explicit inference budgets, would help separate interface effects from model and protocol differences.

\subsection{Threshold Selection and Deployment Precision}
\label{sec:deployment-precision}

The gains from threshold selection on the benchmark (RQ3), contrasted with their limited transfer to reconstructed workflows (RQ7), suggest that thresholds should be selected according to the deployment setting and the consequences of classification errors. Maximizing F1 on a balanced benchmark need not produce an acceptable alert volume in practice, and our frozen global thresholds lowered external F1. Likewise, scores between zero and one should not be interpreted as probabilities without calibration evidence; an F1-optimal threshold below 0.5 does not itself establish miscalibration.

Smell prevalence is particularly important when interpreting precision. At threshold 0.5, \jev{} achieves 72.51\% recall and a 9.47\% false-positive rate. If these rates remained unchanged in a population with 10\% positive workflow--smell pairs, approximately 46\% of its alerts would be correct, compared with 88.45\% on the balanced benchmark. This hypothetical extrapolation illustrates how rarer positives can substantially reduce precision even when recall and false-positive rates remain unchanged. The historical comparison (RQ7) illustrates that these rates can change: at 31.5\% prevalence, main-benchmark rates predict 77.9\% precision for \jev{}, but its false-positive rate rises to 14.86\% and observed precision is 71.7\%. Deployment estimates therefore require representative data.

We therefore recommend evaluating thresholds and review policies on independently labeled workflows representative of the target population. Evaluation should measure both missed smells and the resulting review workload, including whether deferring uncertain predictions concentrates errors sufficiently to justify human inspection. Although our results motivate such policies, they do not establish deployment-ready thresholds or calibrated uncertainty estimates.

\subsection{Repeatability Is Not Robustness}
\label{sec:repeatability-robustness}

RQ5 and RQ6 show that repeatability on identical inputs and robustness to edits that preserve the assessed smells are distinct properties. In continuous workflow analysis, a warning that appears or disappears after adding a neutral comment creates inconsistent feedback, even when aggregate F1 remains unchanged. We thus recommend measuring individual prediction changes under such edits and comparing them with a same-input repetition baseline.

The score changes observed in RQ6 also have implications for review policies. Near-threshold review may capture some instability, but cannot address the large score reversals observed for \gptdec{}. Robustness checks should therefore include predictions far from the threshold rather than treating extreme scores as evidence of invariance. However, for \gptdec{}, we cannot exclude a provider-side update between the original and transformed run as an alternative explanation for the observed changes. These results motivate evaluating both the consistency of individual diagnostics and aggregate detection effectiveness.

\subsection{Implications}
\label{sec:implications}

\noindent\textbf{Support extensible checks with sufficient context.} Our implementation accepts natural-language criteria without a dedicated detection rule for each smell. However, RQ1--RQ7 evaluate an existing catalog; they do not test evolving syntax, new definitions, or reduced authoring effort. The static comparison (RQ7) supports established rules for the evaluated checks, while complementary errors motivate further investigation of model-assisted analysis. To support criteria that depend on information beyond the supplied YAML, specifications should identify required context, such as repository settings or called workflows; tools should obtain it or expose its absence.

\noindent\textbf{Validate routing and filtering policies.} The complementary errors observed in RQ7 suggest that decision models could help review selected static-analysis warnings, but the resulting filtering policies remain unevaluated. Before deployment, the routing policy should be fixed and evaluated on held-out workflows, measuring both rejected false alarms and suppressed true positives. These measurements should determine whether the reduction in review workload justifies the missed findings.

\noindent\textbf{Explore interactive IDE feedback.} The latency and API cost measured in RQ4 motivate IDE integrations that flag potential workflow smells during editing, similarly to static-analysis diagnostics. Prior work on live feedback for behavioral changes illustrates this integration direction and highlights challenges in deciding when to analyze edits and how to present actionable feedback~\cite{soares2013live}. Such integration requires evaluating end-to-end responsiveness, handling incomplete YAML, and managing warnings that change after irrelevant edits. Developer studies should establish whether this feedback improves issue detection without excessive interruptions or review effort~\cite{johnson2013static,sadowski2015tricorder}.

\noindent\textbf{Make judgments inspectable.} Each finding should expose the assessed criterion, the relevant workflow evidence, and any missing context needed to interpret the verdict. Generated explanations should be checked for consistency with the verdict; matching an evidence excerpt to the YAML establishes its origin, not its interpretation.

\noindent\textbf{Match deployment to data constraints.} Hosted inference requires transmitting workflow contents to an external service. Local execution may be preferable when organizational policies prohibit this transfer, but model selection must also consider detection effectiveness, latency, and hardware and maintenance costs.

\section{Threats to Validity}
\label{sec:threats}

This section discusses threats to the validity of our findings.

\noindent\textbf{Construct validity.} The 19 smells and their operational definitions derive from prior studies~\cite{uchoa2026gha,khatami2024catching}. Khatami et al.\ confirmed only 7 of their 22 candidate smells through maintainer responses; on the corresponding seven smells, micro-F1 is 90.69\%, 83.92\%, and 42.91\% for \jev{}, \gptdec{}, and \tevone{}, versus 71.22\%, 59.54\%, and 55.44\% on the other 12. Manual annotations support evaluation against these criteria but do not establish that maintainers consider every finding actionable. Some judgments also require repository settings, reusable workflows, or project intent unavailable in the supplied YAML. We evaluate only annotated workflow--smell pairs; predictions for other criteria remain unvalidated. The balanced benchmark does not establish deployment precision or alert volume (Section~\ref{sec:deployment-precision}). Generated explanations were not evaluated for correctness or faithfulness.

\noindent\textbf{Internal validity.} We reuse annotations produced by six annotators working in three pairs, with disagreements resolved through discussion. Pre-consensus agreement ($\kappa=0.848$) does not exclude shared interpretation errors. Candidate labels originated from the static detector, and only candidates confirmed by the annotators were retained. The released detector reproduces 1,709 reference labels, while one pair causes a parser error. This near-complete agreement reflects the benchmark's construction, which excludes detector--annotator disagreements; these may be among the harder cases, so results may overstate effectiveness on unfiltered workflows. The benchmark therefore cannot establish an advantage over that detector. RQ7 provides a separate comparison with historical manual assessments.

In addition, 45 of the 50 positive \smell{UnsafeDeployOnFork} examples were constructed from real workflows, potentially introducing recognizable regularities. The 903 workflows also contain 872 distinct YAML files; 16 duplicate groups span repositories and cross-validation folds, assigning repositories linked by identical files to the same fold leaves cross-validated micro-F1 unchanged. Prior inspection of the benchmark may also have influenced specifications and prompts.

The configurations differ in training, prompts, batching, response policies, and deployment. Their comparison measures complete inference configurations rather than isolating model size or interface effects. Adapter behavior is another potential source of error. All \gemmafour{} responses passed structural validation. For the examined zero-F1 smells, \gptdec{}'s high AUROC is consistent with threshold effects, although adapter correctness was not independently established. \tevone{}'s inverted rankings remain unexplained.

\noindent\textbf{External validity.} The 903 workflows from 559 open-source repositories were not randomly sampled. Their size (7--207 YAML lines; median 45) and public development context limit generalization to larger workflows, private repositories, and enterprise CI/CD. Public workflows may have appeared in training data~\cite{sallou2024breaking}. Metamorphic testing does not rule out contamination because transformed inputs retain substantial similarity to the originals. We also did not assess adversarial workflow text, which could influence judgments through indirect prompt injection~\cite{greshake2023injection}.
The static-analysis comparison uses 46 reconstructed workflows from 23 repositories, with few positive instances for several smells. Revisions and column mappings were recovered using recorded detector outputs, without consulting manual labels; matching those outputs does not guarantee recovery of the original commits and may favor the detector. The historical assessments were also produced by one assessor for that detector rather than our specifications, so some disagreements may reflect differences in criteria. Independent annotation of a broader workflow sample is needed to validate the comparison.

\noindent\textbf{Conclusion validity.} F1 does not capture differing consequences of false alarms and missed smells. Repository-cluster bootstrap intervals and paired permutation tests account for within-project dependence but remain conditional on the sampled repositories and resampling assumptions. Cross-validation separates threshold selection from scoring, while providing no independent validation of specifications, prompts, model selection, or deployment prevalence. The exploratory hybrid policy was selected after inspecting reference labels and therefore does not estimate prospective hybrid performance.
Three original-input runs provide limited evidence about repeatability; agreement across them does not establish determinism. Each generative configuration and each transformed-input evaluation was executed once. Thus, confidence intervals based on resampling pairs or repositories do not capture run-to-run variation. Original and transformed \gptdec{} runs occurred about 26 hours apart, leaving provider-side changes as a possible confound. Each workflow received one transformation, preventing attribution of differences between transformation groups solely to the operators. Robustness intervals concern aggregate F1 differences, not flip rates. Comparisons involving \luna{} must also account for its incomplete coverage and potentially different classified subsets.

\noindent\textbf{Operational validity and reproducibility.} Measurements reflect the model versions and settings used in October 2026, including the Mac Mini M4 Pro with 64 GB of unified memory used locally. Different batching arrangements and generated outputs limit per-request comparisons; token counts depend on provider conventions, and summed request durations are not end-to-end execution times. Local execution avoids API charges, but electricity, hardware, and maintenance costs were not measured. The projected full-catalog cost of \luna{} assumes unchanged average usage per pair and is not an observed run.
Increasing \tevone{}'s context limit enabled execution but does not establish reliable interpretation of inputs longer than those seen during fine-tuning. For 57 workflows, reported input usage divided by 22 exceeds 1,500 tokens, approximately the longest reported training input~\cite{together2026tev1}; individual question lengths were not recorded separately. Hosted-service updates and incompletely recorded local model metadata, including \gemmafour{}'s exact digest and quantization, may limit exact reproduction. Finally, the study does not measure specification-authoring effort, developer review effort, or the usefulness of interactive IDE feedback; these remain proposed applications requiring evaluation.

\section{Related Work}
\label{sec:related}

We review related work on CI/CD quality, declarative smell detection, and model-based analysis.

\subsection{Quality of GitHub Actions Workflows}

Studies of GitHub Actions adoption~\cite{decan2022use} and workflow evolution~\cite{cardoen2024dataset} establish workflows as software artifacts requiring maintenance. Their combination of triggers, permissions, external actions, and execution environments creates configuration concerns involving security, reliability, and resource consumption~\cite{koishybayev2022characterizing,muralee2023argus,bouzenia2024resource}.

Research on pipeline quality predates GitHub Actions. Gallaba and McIntosh investigated misuse of Travis CI features~\cite{gallaba2020misuse}, Zampetti et al.\ derived CI bad practices from developer discussions~\cite{zampetti2020badpractices}, and Vassallo et al.\ proposed CD-Linter for GitLab CI/CD smells and investigated developer reactions to its warnings~\cite{vassallo2020cdlinter}. For GitHub Actions, studies have examined resource consumption~\cite{bouzenia2024resource}, automation maintenance costs~\cite{valenzuela2024hidden}, and security practices and misconfigurations~\cite{huang2025revisiting,riggio2025pipelines}. Silva et al.~\cite{silva2023factors} investigated factors associated with build failure correction time in CI, while Silva et al.~\cite{silva2025unveiling} examined the relationship between CI and code review in ten closed-source projects. These studies highlight the implications of CI practices for development activities, complementing research on workflow configuration quality.

Khatami et al.\ proposed a catalog of workflow smells, implemented a rule-based detector, and investigated its effectiveness and maintainer responses to proposed fixes~\cite{khatami2024catching}. Our catalog and benchmark build on this work through the earlier study~\cite{uchoa2026gha}. We investigate whether models can apply these criteria through natural-language specifications and compare decision-oriented inference with the rule-based detector using historical manual assessments.

\subsection{Smells in Infrastructure as Code}

Sharma et al.\ cataloged implementation and design smells in Puppet code~\cite{sharma2016configsmells}, while Rahman et al.\ identified seven categories of security smells and implemented the SLIC static analyzer~\cite{rahman2019sevensins}. Bessghaier et al.\ examined associations between IaC smells, change frequency, and defect proneness~\cite{bessghaier2024iacsmells}. These studies provide foundations for investigating configuration quality, although GitHub Actions introduces distinct criteria involving event triggers, permission inheritance, external actions, and interactions among jobs.

Masoumzadeh et al.\ reported substantial disagreement among inspectors assessing IaC smells~\cite{masoumzadeh2025agreement}. Such findings highlight the importance of explicit applicability conditions and exceptions when constructing reference labels. Our evaluation uses operational definitions and manual assessments, while acknowledging that unavailable project context and differing interpretations can affect judgments.

\subsection{Static Analysis and Declarative Smell Detection}

DECOR describes code and design smells in a high-level specification language from which detection algorithms are generated~\cite{moha2010decor}. Our approach similarly separates specifications from detection, but uses models to interpret natural-language criteria. GLITCH detects IaC smells through a technology-independent intermediate representation~\cite{saavedra2022glitch,saavedra2023polyglot}, demonstrating another route to reusable analysis across technologies. Data-flow analysis can additionally capture relationships relevant to IaC security smells~\cite{opdebeeck2023dataflow}.

For GitHub Actions, GASH detects workflow smells~\cite{freitas2024gash}, GHAST supports security assessment~\cite{benedetti2022automatic}, actionlint validates workflow syntax and semantics~\cite{actionlint}, and zizmor detects security-relevant patterns~\cite{zizmor}. Koishybayev et al.\ characterized workflow weaknesses~\cite{koishybayev2022characterizing}, while ARGUS applies taint analysis to injection paths from untrusted inputs~\cite{muralee2023argus}. Beyond automated detection, de Mello et al.~\cite{de2022recommendations} investigated the challenges developers face when identifying code smells, emphasizing the importance of examining the surrounding context. This perspective motivates evaluating whether model-based detectors can consistently interpret applicability conditions and exceptions, rather than relying solely on isolated configuration patterns.

These approaches differ in how checks are expressed and evaluated. Static analyzers encode supported properties through rules or semantic analyses, whereas our approach supplies natural-language criteria to model-specific adapters. Comparisons nevertheless require aligned criteria and reference annotations. Our empirical baseline therefore uses Khatami et al.'s detector~\cite{khatami2024catching}, whose catalog underlies the evaluated criteria. Comparing other tools requires aligning their checks and exceptions with these criteria.

\subsection{Language Models for Smell Detection and CI/CD Analysis}

Silva et al.\ evaluated ChatGPT for detecting Java code smells~\cite{silva2024chatgpt}, Souza et al.\ compared language models with static analysis for code smell detection~\cite{souza2026beyond}, and Lucas et al.\ investigated test smell detection~\cite{lucas2024evaluating}. For IaC, SecLLM applies tailored prompting to security smell detection~\cite{devito2025secllm}, while Vo et al.\ studied prompted and fine-tuned models for Terraform smells~\cite{vo2025terraform,vo2026finetuned}. These studies investigate model-based inspection across code and configuration artifacts.

Within GitHub Actions research, Ghaleb and Rathnayake investigated workflow generation from natural-language requirements~\cite{ghaleb2025llmci}. More closely related, Abrokwah and Ghaleb evaluated a documentation-grounded checklist of 30 compliance criteria using open-weight models, resolving model disagreements through model-based adjudication and manual review of a subset~\cite{abrokwah2026compliant}. Both their study and ours apply explicit criteria to existing workflows. Our evaluation emphasizes decision-oriented inference, threshold and abstention policies, and the distinction between repeated-input agreement and robustness under label-preserving transformations. Differences in criteria and reference assessments prevent direct comparison of headline effectiveness.

\subsection{Decision Models and Selective Classification}

Decision interfaces evaluate inputs against bounded questions and return answers or scores. Geuter et al.\ investigated decision-oriented readout and fine-tuning for open-weight language models~\cite{geuter2026inner}. Yang et al.\ examined limitations of one-pass decision models on tasks requiring intermediate reasoning~\cite{yang2026limits}, while Tang and Zheng's early evidence audit proposed evaluation guidance emphasizing calibration, selective control, and matched baselines~\cite{tang2026tdm}. Recent preprints also report sensitivity to option names~\cite{sun2026typesafe} and optimized context additions~\cite{xu2026jevout}, and examine calibration under prevalence shift~\cite{iqbal2026prevalence}. These studies motivate examining how decision-oriented inference performs on workflow criteria requiring different kinds of evidence.

Calibration concerns the correspondence between predicted probabilities and observed outcomes~\cite{guo2017calibration}, while selective classification trades coverage for reliability by allowing abstention~\cite{geifman2017selective}. We examine related operational choices through threshold sensitivity, repository-grouped threshold selection, and explicit accounting for uncertain responses and failed requests. These analyses characterize decision policies without establishing that the returned scores are calibrated probabilities.

\subsection{Repeatability and Robustness of Model-Based Analyzers}

Language-model outputs can vary across identical inputs even under nominally deterministic settings~\cite{atil2025nondeterminismdeterministicllmsettings}. Metamorphic testing checks expected relations between outputs for related inputs~\cite{metamorphic-testing-2,segura2016survey}; Applis et al.\ applied semantics-preserving transformations to models of code~\cite{metamorphic-testing}. We adapt this approach to GitHub Actions using transformations that preserve the assessed smell labels, comparing prediction changes with an identical-input repetition baseline. This evaluates robustness beyond aggregate effectiveness, while leaving training-data contamination as a separate validity concern~\cite{sallou2024breaking}.

\section{Conclusion}
\label{sec:conclusion}

We evaluated definition-driven GitHub Actions smell detection using five model configurations on 1,710 annotated pairs from 903 workflows. Decision models apply reusable natural-language criteria with effectiveness that varies by smell and decision policy. \jev{} reaches 85.43\% micro-F1 with cross-validated thresholds, while its median request latency of 0.675 seconds and approximately US\$0.22 in benchmark API charges motivate frequent workflow screening. Local \tevone{} achieves lower effectiveness and higher median latency in our environment, but local inference may suit policies that prohibit external processing.

Decision models are a low-cost candidate for definition-driven checks. They evaluate every criterion for a workflow in a single low-latency request, cover all pairs, are highly repeatable, and expose scores that support adjustable decision policies. Because each criterion is a natural-language definition rather than a dedicated detection rule, they make it possible to prototype and refine new checks quickly, although the resulting reduction in authoring effort remains to be measured. Three findings qualify their use. Global thresholds selected on the benchmark transfer poorly to the historical comparison set. Repeatability does not ensure robustness: two decision models reproduce identical labels across three runs yet change more than 3\% of their predictions after smell-preserving transformations. Finally, the static analyzer performs best on the 46 reconstructed workflows, especially for localized criteria, and model-based filtering rejects some false alarms while also suppressing true positives. The observed costs and complementary errors motivate further evaluation of decision models for prototyping checks and assisting static analysis, with policies validated on independent data.

The ability to express criteria as specifications supports prototyping additional checks, while effectiveness on new definitions and reductions in authoring effort remain untested. Future work should use independently annotated workflows, validate threshold and filtering policies, and assess sensitivity to specification wording. Developer studies should establish whether these capabilities improve diagnostics and reduce review effort.

\section*{Data Availability}

The data and replication materials are available~\cite{artefatos}.

\section*{Acknowledgment}

This work was partially supported by CNPq, FAPESQ-PB, FUNCAP, and CAPES. 

\bibliographystyle{IEEEtran}
\bibliography{references}

@incollection{basili1994gqm,
  author    = {Victor R. Basili and Gianluigi Caldiera and H. Dieter Rombach},
  title     = {The Goal Question Metric Approach},
  booktitle = {Encyclopedia of Software Engineering},
  editor    = {John J. Marciniak},
  publisher = {John Wiley \& Sons},
  volume    = {1},
  pages     = {528--532},
  year      = {1994},
}

@inproceedings{khatami2024catching,
  author    = {Ali Khatami and C{\'e}dric Willekens and Andy Zaidman},
  title     = {Catching Smells in the Act: {A} {GitHub} {Actions} Workflow Investigation},
  booktitle = {2024 IEEE International Conference on Source Code Analysis and Manipulation (SCAM)},
  pages     = {47--58},
  publisher = {IEEE},
  year      = {2024},
  doi       = {10.1109/SCAM63643.2024.00015}
}

@inproceedings{freitas2024gash,
  author    = {Matheus B. Freitas and Lincoln S. Rocha},
  title     = {{GASH}: The {GitHub} {Actions} Smell Hunter},
  booktitle = {Workshop de Visualiza{\c{c}}{\~a}o, Evolu{\c{c}}{\~a}o e Manuten{\c{c}}{\~a}o de Software (VEM)},
  pages     = {103--112},
  year      = {2024},
  doi       = {10.5753/vem.2024.3911}
}

@inproceedings{benedetti2022automatic,
  author    = {Giacomo Benedetti and Luca Verderame and Alessio Merlo},
  title     = {Automatic Security Assessment of {GitHub} {Actions} Workflows},
  booktitle = {Proceedings of the 2022 ACM Workshop on Software Supply Chain Offensive Research and Ecosystem Defenses},
  pages     = {37--45},
  year      = {2022},
  publisher = {ACM}
}

@inproceedings{koishybayev2022characterizing,
  author    = {Igibek Koishybayev and Aleksandr Nahapetyan and Raima Zachariah and Siddharth Muralee and Bradley Reaves and Alexandros Kapravelos and Aravind Machiry},
  title     = {Characterizing the Security of {GitHub} {CI} Workflows},
  booktitle = {31st USENIX Security Symposium (USENIX Security 22)},
  pages     = {2747--2763},
  year      = {2022},
  publisher = {USENIX Association},
}

@inproceedings{bouzenia2024resource,
  author    = {Islem Bouzenia and Michael Pradel},
  title     = {Resource Usage and Optimization Opportunities in Workflows of {GitHub} {Actions}},
  booktitle = {Proceedings of the 46th IEEE/ACM International Conference on Software Engineering (ICSE)},
  publisher = {ACM},
  year      = {2024},
  doi       = {10.1145/3597503.3623303},
}

@inproceedings{valenzuela2024hidden,
  author    = {Pablo Valenzuela-Toledo and Alexandre Bergel and Timo Kehrer and Oscar Nierstrasz},
  title     = {The Hidden Costs of Automation: {An} Empirical Study on {GitHub} {Actions} Workflow Maintenance},
  booktitle = {2024 IEEE International Conference on Source Code Analysis and Manipulation (SCAM)},
  pages     = {213--223},
  year      = {2024},
  publisher = {IEEE}
}

@inproceedings{cardoen2024dataset,
  author    = {Guillaume Cardoen and Tom Mens and Alexandre Decan},
  title     = {A Dataset of {GitHub} {Actions} Workflow Histories},
  booktitle = {Proceedings of the 21st International Conference on Mining Software Repositories (MSR)},
  pages     = {677--681},
  year      = {2024},
  publisher = {ACM},
  doi       = {10.1145/3643991.3644867}
}

@misc{typesafe2026api,
  author       = {{TypeSafe AI}},
  title        = {{System One} {API} Reference},
  year         = {2026},
  howpublished = {\url{https://docs.typesafe.ai/api/system-one}},
  note         = {Product documentation; accessed 2026-10-27}
}

@misc{typesafe2026models,
  author       = {{TypeSafe AI}},
  title        = {Models and Pricing},
  year         = {2026},
  howpublished = {\url{https://docs.typesafe.ai/models}},
  note         = {Dated pricing reference; accessed 2026-10-27}
}

@inproceedings{guo2017calibration,
  author    = {Chuan Guo and Geoff Pleiss and Yu Sun and Kilian Q. Weinberger},
  title     = {On Calibration of Modern Neural Networks},
  booktitle = {Proceedings of the 34th International Conference on Machine Learning (ICML)},
  pages     = {1321--1330},
  year      = {2017},
  publisher = {PMLR}
}

@inproceedings{geifman2017selective,
  author    = {Yonatan Geifman and Ran El-Yaniv},
  title     = {Selective Classification for Deep Neural Networks},
  booktitle = {Advances in Neural Information Processing Systems 30 (NeurIPS)},
  year      = {2017},
}

@inproceedings{greshake2023injection,
  author    = {Kai Greshake and Sahar Abdelnabi and Shailesh Mishra and Christoph Endres and Thorsten Holz and Mario Fritz},
  title     = {Not What You've Signed Up For: Compromising Real-World {LLM}-Integrated Applications with Indirect Prompt Injection},
  booktitle = {Proceedings of the 16th ACM Workshop on Artificial Intelligence and Security (AISec)},
  year      = {2023},
  doi       = {10.1145/3605764.3623985}
}

@misc{github2026syntax,
  author       = {{GitHub}},
  title        = {Workflow Syntax for {GitHub} {Actions}},
  year         = {2026},
  howpublished = {\url{https://docs.github.com/en/actions/reference/workflows-and-actions/workflow-syntax}},
  note         = {Official documentation; accessed 2026-10-27}
}

@inproceedings{lucas2024evaluating,
  author    = {Keila Lucas and Rohit Gheyi and Elvys Soares and M{\'a}rcio Ribeiro and Ivan Machado},
  title     = {Evaluating Large Language Models in Detecting Test Smells},
  booktitle = {Proceedings of the Brazilian Symposium on Software Engineering (SBES), Insightful Ideas and Emerging Results Track},
  year      = {2024},
  note      = {arXiv:2407.19261}
}

@inproceedings{decan2022use,
  author    = {Alexandre Decan and Tom Mens and Pooja Rani Mazrae and Mehdi Golzadeh},
  title     = {On the Use of {GitHub} {Actions} in Software Development Repositories},
  booktitle = {Proceedings of the IEEE International Conference on Software Maintenance and Evolution (ICSME)},
  year      = {2022},
  publisher = {IEEE}
}

@inproceedings{muralee2023argus,
  author    = {Siddharth Muralee and Igibek Koishybayev and Aleksandr Nahapetyan and Greg Tystahl and Brad Reaves and Antonio Bianchi and William Enck and Alexandros Kapravelos and Aravind Machiry},
  title     = {{ARGUS}: A Framework for Staged Static Taint Analysis of {GitHub} Workflows and Actions},
  booktitle = {32nd USENIX Security Symposium (USENIX Security 23)},
  pages     = {6983--7000},
  year      = {2023},
  publisher = {USENIX Association}
}

@inproceedings{ghaleb2025llmci,
  author    = {Taher A. Ghaleb and Dulina Rathnayake},
  title     = {Can {LLMs} Write {CI}? {A} Study on Automatic Generation of {GitHub} {Actions} Configurations},
  booktitle = {Proceedings of the IEEE International Conference on Software Maintenance and Evolution (ICSME), NIER Track},
  year      = {2025},
  note      = {arXiv:2507.17165}
}

@inproceedings{abrokwah2026compliant,
  author    = {Edward Abrokwah and Taher A. Ghaleb},
  title     = {How Compliant Are {GitHub} {Actions} Workflows? {A} Checklist-Based Study with {LLM}-Assisted Auditing},
  booktitle = {Proceedings of the 30th International Conference on Evaluation and Assessment in Software Engineering (EASE)},
  year      = {2026},
  note      = {arXiv:2605.02091}
}

@misc{actionlint,
  author       = {{rhysd}},
  title        = {actionlint: Static Checker for {GitHub} {Actions} Workflow Files},
  howpublished = {\url{https://github.com/rhysd/actionlint}},
  note         = {Accessed 2026-10-27}
}

@misc{zizmor,
  author       = {William Woodruff and others},
  title        = {zizmor: Static Analysis for {GitHub} {Actions}},
  howpublished = {\url{https://github.com/zizmorcore/zizmor}},
  note         = {Accessed 2026-10-27}
}

@misc{openai2026decisions,
  author       = {{OpenAI}},
  title        = {{Decisions API} Guide},
  year         = {2026},
  howpublished = {\url{https://developers.openai.com/api/docs/guides/decisions}},
  note         = {Accessed: October 2026}
}

@book{efron1993bootstrap,
  author    = {Bradley Efron and Robert J. Tibshirani},
  title     = {An Introduction to the Bootstrap},
  publisher = {Chapman \& Hall},
  year      = {1993}
}

@book{davison1997bootstrap,
  author    = {A. C. Davison and D. V. Hinkley},
  title     = {Bootstrap Methods and Their Application},
  publisher = {Cambridge University Press},
  year      = {1997}
}

@book{good2005permutation,
  author    = {Phillip I. Good},
  title     = {Permutation, Parametric, and Bootstrap Tests
               of Hypotheses},
  edition   = {3},
  publisher = {Springer},
  year      = {2005},
  doi       = {10.1007/b138696}
}

@article{field2007bootstrapping,
  author  = {C. A. Field and A. H. Welsh},
  title   = {Bootstrapping Clustered Data},
  journal = {Journal of the Royal Statistical Society:
             Series B (Statistical Methodology)},
  volume  = {69},
  number  = {3},
  pages   = {369--390},
  year    = {2007},
  doi     = {10.1111/j.1467-9868.2007.00593.x}
}

@article{holm1979simple,
  author  = {Sture Holm},
  title   = {A Simple Sequentially Rejective
             Multiple Test Procedure},
  journal = {Scandinavian Journal of Statistics},
  volume  = {6},
  number  = {2},
  pages   = {65--70},
  year    = {1979}
}

@article{moha2010decor,
  author  = {Naouel Moha and Yann-Ga{\"e}l Gu{\'e}h{\'e}neuc and Laurence Duchien and Anne-Fran{\c{c}}oise Le Meur},
  title   = {{DECOR}: {A} Method for the Specification and Detection of Code and Design Smells},
  journal = {IEEE Transactions on Software Engineering},
  volume  = {36},
  number  = {1},
  pages   = {20--36},
  year    = {2010},
  doi     = {10.1109/TSE.2009.50}
}

@inproceedings{sharma2016configsmells,
  author    = {Tushar Sharma and Marios Fragkoulis and Diomidis Spinellis},
  title     = {Does Your Configuration Code Smell?},
  booktitle = {Proceedings of the 13th International Conference on Mining Software Repositories},
  pages     = {189--200},
  year      = {2016},
  publisher = {ACM},
  doi       = {10.1145/2901739.2901761}
}

@inproceedings{rahman2019sevensins,
  author    = {Akond Rahman and Chris Parnin and Laurie Williams},
  title     = {The Seven Sins: {Security} Smells in Infrastructure as Code Scripts},
  booktitle = {Proceedings of the 41st International Conference on Software Engineering},
  pages     = {164--175},
  year      = {2019},
  doi       = {10.1109/ICSE.2019.00033}
}

@inproceedings{saavedra2022glitch,
  author    = {Nuno Saavedra and Jo{\~a}o F. Ferreira},
  title     = {{GLITCH}: {Automated} Polyglot Security Smell Detection in Infrastructure as Code},
  booktitle = {Proceedings of the 37th IEEE/ACM International Conference on Automated Software Engineering},
  year      = {2022},
  publisher = {ACM},
  doi       = {10.1145/3551349.3556945}
}

@inproceedings{saavedra2023polyglot,
  author    = {Nuno Saavedra and Jo{\~a}o Gon{\c{c}}alves and Miguel Henriques and Jo{\~a}o F. Ferreira and Alexandra Mendes},
  title     = {Polyglot Code Smell Detection for Infrastructure as Code with {GLITCH}},
  booktitle = {Proceedings of the 38th IEEE/ACM International Conference on Automated Software Engineering},
  year      = {2023},
  doi       = {10.1109/ASE56229.2023.00162},
  pages = {2042--2045}
}

@inproceedings{opdebeeck2023dataflow,
  author    = {Ruben Opdebeeck and Ahmed Zerouali and Coen De Roover},
  title     = {Control and Data Flow in Security Smell Detection for Infrastructure as Code: {Is} It Worth the Effort?},
  booktitle = {Proceedings of the 20th International Conference on Mining Software Repositories},
  pages     = {534--545},
  year      = {2023},
  doi       = {10.1109/MSR59073.2023.00079}
}

@inproceedings{bessghaier2024iacsmells,
  author    = {Narjes Bessghaier and Mahi Begoug and Chemseddine Mebarki and Ali Ouni and Mohammed Sayagh and Mohamed Wiem Mkaouer},
  title     = {On the Prevalence, Co-occurrence, and Impact of Infrastructure-as-Code Smells},
  booktitle = {Proceedings of the IEEE International Conference on Software Analysis, Evolution and Reengineering},
  pages     = {23--34},
  year      = {2024},
  doi       = {10.1109/SANER60148.2024.00009}
}

@article{masoumzadeh2025agreement,
  author  = {Sogol Masoumzadeh and Nuno Saavedra and Rungroj Maipradit and Lili Wei and Jo{\~a}o F. Ferreira and D{\'a}niel Varr{\'o} and Shane McIntosh},
  title   = {Do Experts Agree About Smelly Infrastructure?},
  journal = {IEEE Transactions on Software Engineering},
  volume  = {51},
  number  = {5},
  pages   = {1472--1486},
  year    = {2025},
  doi     = {10.1109/TSE.2025.3553383}
}

@article{devito2025secllm,
  author  = {Gabriele De Vito and Fabio Palomba and Filomena Ferrucci},
  title   = {{SecLLM}: {Enhancing} Security Smell Detection in {IaC} With Large Language Models},
  journal = {IEEE Access},
  volume  = {13},
  pages   = {204480--204498},
  year    = {2025},
  doi     = {10.1109/ACCESS.2025.3637505}
}

@inproceedings{vo2025terraform,
  author    = {Quoc-Huy Vo and Ha Dao and Kensuke Fukuda},
  title     = {Harnessing the Power of {LLMs} for Code Smell Detection in {Terraform} Infrastructure as Code},
  booktitle = {Proceedings of the IEEE 49th Annual Computers, Software, and Applications Conference},
  pages     = {533--542},
  year      = {2025},
  doi       = {10.1109/COMPSAC65507.2025.00075}
}

@article{vo2026finetuned,
  author  = {Quoc-Huy Vo and Ha Dao and Kensuke Fukuda},
  title   = {Securing {IaC}: {Detecting} {Terraform} Security Smells with Fine-tuned {LLMs}},
  journal = {Journal of Information Processing},
  volume  = {34},
  pages   = {886--897},
  year    = {2026},
  doi     = {10.2197/ipsjjip.34.886}
}

@inproceedings{silva2024chatgpt,
  author    = {Luciana L. Silva and Janio R. Silva and Jo{\~a}o Eduardo Montandon and Marcus Andrade and Marco Tulio Valente},
  title     = {Detecting Code Smells using {ChatGPT}: {Initial} Insights},
  booktitle = {Proceedings of the 18th ACM/IEEE International Symposium on Empirical Software Engineering and Measurement},
  pages     = {400--406},
  year      = {2024},
  doi       = {10.1145/3674805.3690742}
}

@misc{souza2026beyond,
  author        = {Saymon Souza and Amanda Santana and Eduardo Figueiredo and Igor Muzetti and Jo{\~a}o Eduardo Montandon and Lionel Briand},
  title         = {Beyond Strict Rules: {Assessing} the Effectiveness of Large Language Models for Code Smell Detection},
  year          = {2026},
  archivePrefix = {arXiv},
  eprint        = {2601.09873},
  url           = {https://arxiv.org/abs/2601.09873},
}

@misc{geuter2026inner,
  author        = {Jonathan Geuter and Andrew Cheng and David Alvarez-Melis},
  title         = {Releasing the Inner {Jev}: {A} Simple Recipe for Open Decision Models},
  year          = {2026},
  archivePrefix = {arXiv},
  eprint        = {2610.17205},
  url           = {https://arxiv.org/abs/2610.17205},
  note          = {Preprint}
}

@misc{yang2026limits,
  author        = {Wang Yang and Xinpeng Li and Yanyan Zhang and Chuang Ma and Shouren Wang and Jianzhe Wei and Liqin Ye and Xiaoran Cai and Mohsen Hariri and Vipin Chaudhary and Xiaotian Han},
  title         = {What Can't {Jev} Decide? {Probing} the Limits of One-Pass Decision Models},
  year          = {2026},
  archivePrefix = {arXiv},
  eprint        = {2610.24929},
  url           = {https://arxiv.org/abs/2610.24929},
  note          = {Preprint}
}

@inproceedings{huang2025revisiting,
  author    = {Jiangnan Huang and Bin Lin},
  title     = {Revisiting Security Practices for {GitHub} {Actions} Workflows},
  booktitle = {Proceedings of the 33rd IEEE/ACM International Conference on Program Comprehension},
  pages     = {73--77},
  year      = {2025},
  doi       = {10.1109/ICPC66645.2025.00016}
}

@inproceedings{riggio2025pipelines,
  author    = {Edoardo Riggio and Cesare Pautasso},
  title     = {Pipelines Under Pressure: {An} Empirical Study of Security Misconfigurations of {GitHub} Workflows},
  booktitle = {Proceedings of the 26th International Conference on Product-Focused Software Process Improvement},
  pages     = {220--236},
  year      = {2025},
  publisher = {Springer},
  doi       = {10.1007/978-3-032-12089-2_14}
}

@misc{atil2025nondeterminismdeterministicllmsettings,
      title={Non-Determinism of ``Deterministic'' {LLM} Settings}, 
      author={Berk Atil and Sarp Aykent and Alexa Chittams and Lisheng Fu and Rebecca J. Passonneau and Evan Radcliffe and Guru Rajan Rajagopal and Adam Sloan and Tomasz Tudrej and Ferhan Ture and Zhe Wu and Lixinyu Xu and Breck Baldwin},
      year={2025},
      eprint={2408.04667},
      archivePrefix={arXiv},
      primaryClass={cs.CL},
      note={arXiv:2408.04667},
}

@inproceedings{johnson2013static,
  author    = {Brittany Johnson and Yoonki Song and Emerson Murphy-Hill and Robert Bowdidge},
  title     = {Why Don't Software Developers Use Static Analysis Tools to Find Bugs?},
  booktitle = {Proceedings of the 35th International Conference on Software Engineering (ICSE)},
  year      = {2013},
  pages     = {672--681},
  doi       = {10.1109/ICSE.2013.6606613}
}

@inproceedings{sadowski2015tricorder,
  author    = {Caitlin Sadowski and Jeffrey van Gogh and Ciera Jaspan and Emma S{\"o}derberg and Collin Winter},
  title     = {Tricorder: {Building} a Program Analysis Ecosystem},
  booktitle = {Proceedings of the 37th International Conference on Software Engineering (ICSE)},
  year      = {2015},
  pages     = {598--608},
  doi       = {10.1109/ICSE.2015.76}
}

@misc{github2026reusable,
  author       = {{GitHub}},
  title        = {Reusing Workflow Configurations: {Supported} Keywords for Jobs That Call a Reusable Workflow},
  howpublished = {\url{https://docs.github.com/en/actions/reference/workflows-and-actions/reusing-workflow-configurations}},
  note         = {Accessed: October 2026},
  year         = {2026}
}

@inproceedings{vassallo2020cdlinter,
  author    = {Carmine Vassallo and Sebastian Proksch and Anna Jancso and Harald C. Gall and Massimiliano {Di Penta}},
  title     = {Configuration Smells in Continuous Delivery Pipelines: {A} Linter and a Six-Month Study on {GitLab}},
  booktitle = {Proceedings of the 28th ACM Joint European Software Engineering Conference and Symposium on the Foundations of Software Engineering (ESEC/FSE)},
  pages     = {327--337},
  year      = {2020},
  publisher = {ACM},
  doi       = {10.1145/3368089.3409709}
}

@article{zampetti2020badpractices,
  author  = {Fiorella Zampetti and Carmine Vassallo and Sebastiano Panichella and Gerardo Canfora and Harald Gall and Massimiliano {Di Penta}},
  title   = {An Empirical Characterization of Bad Practices in Continuous Integration},
  journal = {Empirical Software Engineering},
  volume  = {25},
  number  = {2},
  pages   = {1095--1135},
  year    = {2020},
  doi     = {10.1007/s10664-019-09785-8}
}

@article{gallaba2020misuse,
  author  = {Keheliya Gallaba and Shane McIntosh},
  title   = {Use and Misuse of Continuous Integration Features: {An} Empirical Study of Projects That (Mis)Use {Travis CI}},
  journal = {IEEE Transactions on Software Engineering},
  volume  = {46},
  number  = {1},
  pages   = {33--50},
  year    = {2020},
  doi     = {10.1109/TSE.2018.2838131}
}

@inproceedings{sallou2024breaking,
  author    = {June Sallou and Thomas Durieux and Annibale Panichella},
  title     = {Breaking the Silence: {The} Threats of Using {LLMs} in Software Engineering},
  booktitle = {Proceedings of the 46th IEEE/ACM International Conference on Software Engineering: New Ideas and Emerging Results (ICSE-NIER)},
  pages     = {102--106},
  year      = {2024},
  publisher = {ACM},
  doi       = {10.1145/3639476.3639764}
}

@misc{tang2026tdm,
  author        = {Lijuan Tang and Yuemeng Zheng},
  title         = {Typed Decision Models: {An} Early Evidence Audit and Evaluation Checklist},
  year          = {2026},
  archivePrefix = {arXiv},
  eprint        = {2609.32160},
  url           = {https://arxiv.org/abs/2609.32160},
  note          = {Preprint}
}

@misc{together2026tev1,
  author       = {{Together AI}},
  title        = {{Tev1}: Model Card and {Ollama} Distribution},
  year         = {2026},
  howpublished = {\url{https://ollama.com/library/tev1}},
  note         = {Accessed: October 2026}
}

@inproceedings{lipton2014thresholding,
  author    = {Zachary C. Lipton and Charles Elkan and Balakrishnan Narayanaswamy},
  title     = {Optimal Thresholding of Classifiers to Maximize {F1} Measure},
  booktitle = {Machine Learning and Knowledge Discovery in Databases (ECML PKDD)},
  series    = {Lecture Notes in Computer Science},
  volume    = {8725},
  pages     = {225--239},
  publisher = {Springer},
  year      = {2014},
  doi       = {10.1007/978-3-662-44851-9_15}
}

@misc{google2026gemma4,
  author       = {{Google DeepMind}},
  title        = {{Gemma 4} Model Card},
  year         = {2026},
  howpublished = {\url{https://ai.google.dev/gemma/docs/core/model_card_4}},
  note         = {Accessed: October 2026}
}

@article{metamorphic-testing-2,
  author       = {Tsong Yueh Chen and
                  Fei{-}Ching Kuo and
                  Huai Liu and
                  Pak{-}Lok Poon and
                  Dave Towey and
                  T. H. Tse and
                  Zhi Quan Zhou},
  title        = {Metamorphic Testing: {A} Review of Challenges and Opportunities},
  journal      = {{ACM} Computing Surveys},
  doi          = {10.1145/3143561},
  volume       = {51},
  number       = {1},
  pages        = {4:1--4:27},
  year         = {2018},
  publisher = {{ACM}},
}

@inproceedings{metamorphic-testing,
  author       = {Leonhard Applis and
                  Annibale Panichella and
                  Arie van Deursen},
  title        = {Assessing Robustness of {ML}-Based Program Analysis Tools using Metamorphic
                  Program Transformations},
  booktitle    = {36th {IEEE/ACM} International Conference on Automated Software Engineering,
                  {ASE} 2021, Melbourne, Australia, November 15-19, 2021},
  pages        = {1377--1381},
  publisher    = {{IEEE}},
  year         = {2021},
}

@article{segura2016survey,
  author  = {Sergio Segura and Gordon Fraser and Ana B. Sanchez and Antonio Ruiz-Cort{\'e}s},
  title   = {A Survey on Metamorphic Testing},
  journal = {IEEE Transactions on Software Engineering},
  volume  = {42},
  number  = {9},
  pages   = {805--824},
  year    = {2016},
  doi     = {10.1109/TSE.2016.2532875}
}

@ARTICLE{rian-agentes-test-smells-2026,
author={Melo, Rian and Simoes, Pedro and Gheyi, Rohit and d'Amorim, Marcelo and Ribeiro, Marcio and Soares, Gustavo and Almeida, Eduardo and Soares, Elvys},
journal={IEEE Software},
title={{Agentic LMs: Hunting Down Test Smells}},
year={2026},
publisher={IEEE Computer Society},
  volume       = {43},
  number       = {1},
  pages        = {32--40},
}

@inproceedings{melo2026reftestmodel,
  author    = {Rian Melo and Rohit Gheyi and Luana Martins and M{\'a}rcio Ribeiro and Fabio Palomba},
  title     = {{RefTestModel}: Catalog-Guided Detection of Test-Specific Refactorings with Foundation Models},
  booktitle = {Proceedings of the 38th IEEE International Conference on Tools with Artificial Intelligence (ICTAI)},
  pages     = {1117--1123},
  year      = {2026},
  publisher = {IEEE}
}

@inproceedings{pedro-sbes-2025,
  author       = {Pedro Simões and
                  Rohit Gheyi and
                  Rian Melo and
                  Jonhnanthan Oliveira and
                  M{\'{a}}rcio Ribeiro and
                  Wesley Assunção},
  title        = {{RefModel}: Detecting Refactorings using Foundation Models},
  booktitle    = {Proceedings of the Brazilian Symposium on Software Engineering},  
  year         = {2025},
  pages = {811--817},
  publisher    = {{SBC}},
}

@inproceedings{soares2013live,
  author    = {Soares, Gustavo and Murphy-Hill, Emerson and Gheyi, Rohit},
  title     = {Live feedback on behavioral changes},
  booktitle = {Proceedings of the 1st International Workshop on Live Programming (LIVE)},
  address   = {San Francisco, CA, USA},
  month     = may,
  year      = {2013},
  pages     = {23--26},
  publisher = {IEEE},
  doi       = {10.1109/LIVE.2013.6617344}
}

@misc{sun2026typesafe,
  author        = {Yu Sun and Junhao Xu and Jiajia Shi and Zijin Yang},
  title         = {Type-Safe Is Not Error-Free: Typed Decision Models Follow the Option Name, Not the Definition Bound to It},
  year          = {2026},
  archivePrefix = {arXiv},
  eprint        = {2609.26758},
  url           = {https://arxiv.org/abs/2609.26758},
  note          = {Preprint}
}

@misc{xu2026jevout,
  author        = {Zixiang Xu and Zirui Song and Chiyu Zhang and Xiuying Chen and Xi Liu and Xiyang Hu and Yue Zhao},
  title         = {{JevOut}: Natural Context Can Flip Decision Models},
  year          = {2026},
  archivePrefix = {arXiv},
  eprint        = {2609.30243},
  url           = {https://arxiv.org/abs/2609.30243},
  note          = {Preprint}
}

@misc{iqbal2026prevalence,
  author        = {Abdul Rahman Iqbal and Muhammad Fawad Nasim and Amir Saleem},
  title         = {Typed Decision Models under Prevalence Shift and Label Budgets: {Decision} Training, Calibration, and Implicit Priors},
  year          = {2026},
  archivePrefix = {arXiv},
  eprint        = {2610.26325},
  url           = {https://arxiv.org/abs/2610.26325},
}

@article{brier1950verification,
  author = {Brier, Glenn W.},
  title = {Verification of Forecasts Expressed in Terms of Probability},
  journal = {Monthly Weather Review},
  year = {1950},
  volume = {78},
  number = {1},
  pages = {1--3},
  doi = {10.1175/1520-0493(1950)078<0001:VOFEIT>2.0.CO;2}
}

@misc{ollama2026,
  author       = {{Ollama}},
  title        = {Ollama: Run Large Language Models Locally},
  year         = {2026},
  howpublished = {\url{https://ollama.com}},
  note         = {Accessed: October 2026}
}

@book{manning2008ir,
  author    = {Christopher D. Manning and Prabhakar Raghavan and Hinrich Sch{\"u}tze},
  title     = {Introduction to Information Retrieval},
  publisher = {Cambridge University Press},
  year      = {2008}
}

@misc{uchoa2026gha,
  title = {{GHA-SmellFix}: Taking Action with Agentic {LLMs} for Detecting and Refactoring {GitHub} {Actions} Workflow Smells},
  author    = {Uch{\^o}a, Anderson and
               Souza, Jhefferson and
               {\'A}vila, Emanuel and
               Silva, Gabriel and
               Lira, Kevin and
               Sun, Hui and
               Ara{\'u}jo, Nelson and
               Cordeiro, Arthur and
               Gheyi, Rohit and
               Assun{\c{c}}{\~a}o, Wesley},
  year      = {2026},
  eprint    = {2611.14963},
  archivePrefix = {arXiv},
  url       = {https://arxiv.org/abs/2611.14963}
}

@misc{artefatos,
  author={Rohit Gheyi and Anderson {Uch{\^o}a} and Wesley Assun{\c{c}}{\~a}o and Francisco Sousa and Baldoino Fonseca and M\'{a}rcio Ribeiro},
  title={Artifacts},
  howpublished={See supplementary material.},
  year={2026}
}

@inproceedings{silva2025unveiling,
  title={Unveiling the Relationship Between Continuous Integration and Code Review: A Study with 10 Closed-Source Projects},
  author={Silva, Ruben and Silva, Publio and Bezerra, Carla and Uch{\^o}a, Anderson and Garcia, Alessandro},
  booktitle={Proceedings of the Brazilian Symposium on Software Engineering (SBES)},
  pages={393--404},
  year={2025},
  publisher={SBC}
}

@article{de2022recommendations,
  title={Recommendations for developers identifying code smells},
  author={de Mello, Rafael and Oliveira, Roberto and Uch{\^o}a, Anderson and Oizumi, Willian and Garcia, Alessandro and Fonseca, Baldoino and de Mello, Fernanda},
  journal={IEEE Software},
  volume={40},
  number={2},
  pages={90--98},
  year={2023},
  publisher={IEEE}
}

@inproceedings{silva2023factors,
  title={What factors affect the build failures correction time? {A} multi-project study},
  author={Silva, Gustavo and Bezerra, Carla and Uch{\^o}a, Anderson and Machado, Ivan},
  booktitle={Proceedings of the 17th Brazilian Symposium on Software Components, Architectures, and Reuse},
  pages={41--50},
  year={2023}
}

@misc{fares2026unpacking,
      title={Unpacking Security Scanners for GitHub Actions Workflows}, 
      author={Madjda Fares and Yogya Gamage and Benoit Baudry},
    year   = {2026},
    eprint = {2601.14455},
    archivePrefix = {arXiv},    
    url           = {https://arxiv.org/abs/2601.14455},    
}
\end{document}